\documentclass[12pt,preprint]{aastex}

\newcommand{\atan}{tan^{-1}}

\newcommand{\psidelay}{$\psi_{delay}$}
\newcommand{\phio}{$\phi_o$}
\newcommand{\chio}{$\chi_o$}

\newcommand{\el}{$\ell$}
\newcommand{\Ptime}{$P_{time}$}

\newcommand{\al}{$\alpha$}
\newcommand{\bt}{$\beta$}
\newcommand{\Pone}{$P_1$}
\newcommand{\Ptwo}{$P_2$}
\newcommand{\Pthree}{$P_3$}

\shorttitle{Quantitative Model of PSR B0943+10}
\shortauthors{Rosen and Clemens}

\begin{document}

\title{A Quantitative Non-radial Oscillation Model for the Subpulses in PSR B0943+10}

\author{R. Rosen \& J. Christopher Clemens}
\affil{Department of Physics and Astronomy, University of North 
Carolina, Chapel Hill, NC 27599-3255}
\email{rrosen@physics.unc.edu, clemens@physics.unc.edu}

\begin{abstract} 

In this paper, we analyze time series measurements of PSR B0943+10 and fit them with a non-radial oscillation model.  The model we apply was first developed for total intensity measurements in an earlier paper, and expanded to encompass linear polarization in a companion paper to this one.  We use PSR B0943+10 for the initial tests of our model because it has a simple geometry, it has been exhaustively studied in the literature, and its behavior is well-documented.  As prelude to quantitative fitting, we have reanalyzed previously published archival data of PSR B0943+10 and uncovered subtle but significant behavior that is difficult to explain in the framework of the drifting spark model.  Our fits of a non-radial oscillation model are able to successfully reproduce the observed behavior in this pulsar.  

\end{abstract}
\keywords{pulsars:individual:PSR B0943+10--pulsars:general---pulsars:polarization---stars:neutron---
stars:oscillations}

\section{Introduction}
\label{intro2}

The regular subpulse behavior and simple morphology of PSR B0943+10 have made this pulsar a fiducial for testing pulsar models, especially those models that incorporate drifting subpulses.  Most drifting subpulse models are based on a generic model proposed by \markcite{rud75}{Ruderman} \& {Sutherland} (1975).  They hypothesized that a vacuum gap forms above the surface of the magnetic polar cap and along the co-rotating magnetosphere.  This vacuum gap results from the depletion of charge due to particle emission from the star.  Because the vacuum gap cannot grow indefinitely, sparks discharge across the gap between the magnetosphere and the stellar surface.  These spark regions are fixed in relation to each other.  Because they occur in the region of the magnetosphere that does not co-rotate with the star, the sparks rotate around the magnetic pole at a period incommensurate with the spin period of the star.  In this model, the drifting subpulses are the observable manifestation of the rotating sparks.  The drifting (or rotating) spark model, as it is commonly known, is the foundation for many current pulsar models \markcite{kom70,bac76}({Komesaroff} 1970; {Backer} 1976).  

\begin{figure}
\begin{center}
\includegraphics{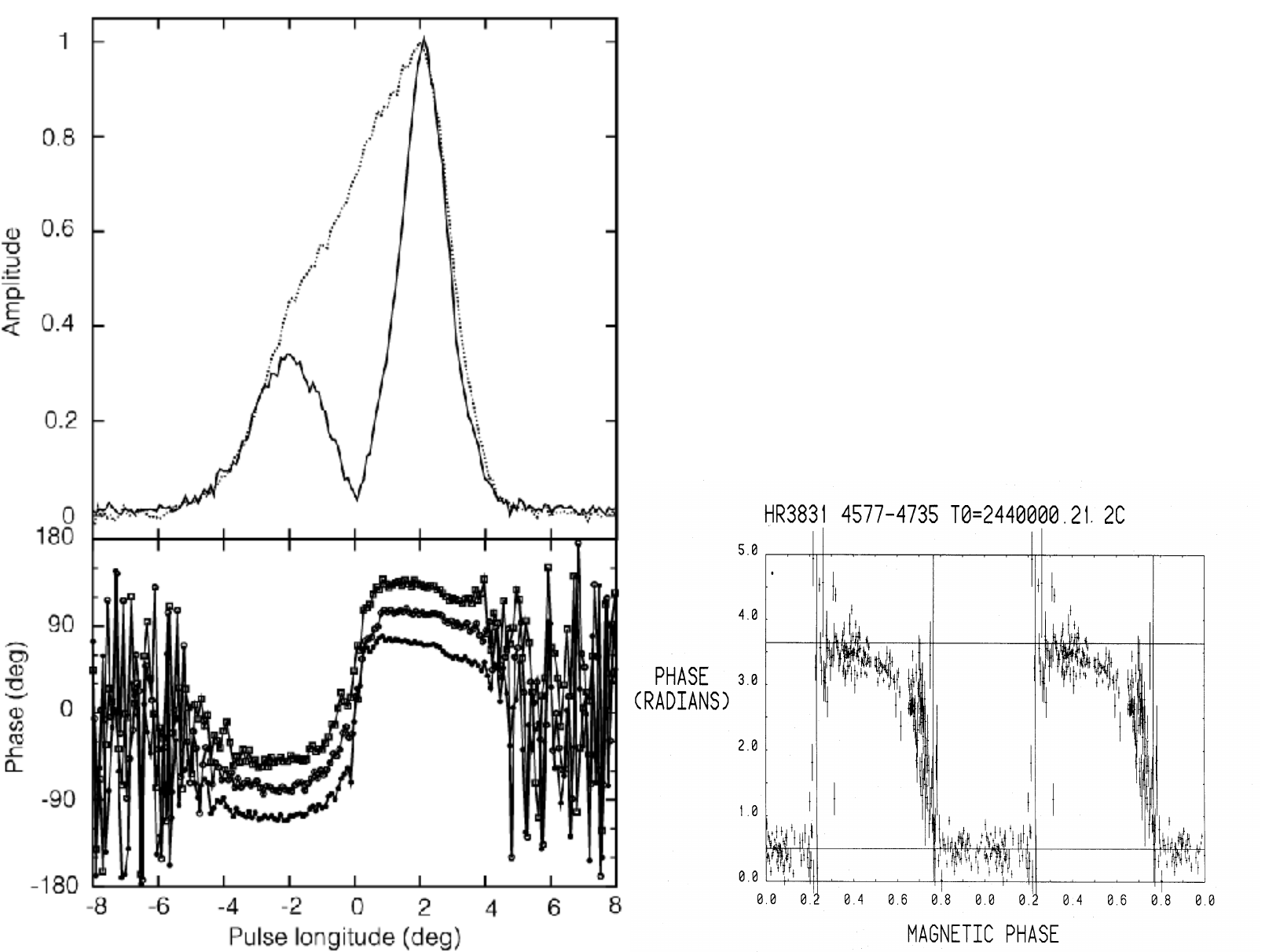}
\end{center}
\caption{Left panel: The average pulse shape, subpulse amplitude (top panel), and phase (lower panel) of PSR B0320+39 \markcite{esl03}({Edwards} {et~al.} 2003).  The subpulse amplitude envelope shows a minimum near zero at the same longitude as a $180^\circ$ shift in the phase.  A $60^{\circ/\circ}$ slope has been removed from the phases.  Right panel: The oscillation phase versus magnetic rotation phase for rapidly oscillating AP star HR 3831 \markcite{kur90}({Kurtz}, {Shibahashi}, \& {Goode} 1990).}
\label{fig:EdwardsKurtz}
\end{figure}

In contrast to the drifting spark model, we have proposed a non-radial pulsation model to explain not only drifting subpulses in pulsars, but also a wide range of other pulsar behavior \markcite{cle04,cle07}({Clemens} \& {Rosen} 2004, 2007).  Pulsations are a widely observed phenomenon in normal stars and in compact objects; white dwarf stars, rapidly oscillating AP stars, delta Scuti stars, and even our sun are known to have oscillation modes.  Non-radial oscillations were previously proposed as an explanation for drifting subpulses \markcite{gol68,vanh80,str92}({Gold} 1968; {van Horn} 1980; {Strohmayer} 1992) but no phenomenological model has been developed to explain the range of behavior seen in pulsars with drifting subpulses.  Our non-radial oscillation model was in its preliminary stages when \markcite{esl03}{Edwards}, {Stappers}, \& {van  Leeuwen} (2003) published their analysis of PSR B0320+39, shown in left panel of Figure \ref{fig:EdwardsKurtz}.  The phase behavior of the subpulses in the data they presented suggested that the modulations we see are a combination of two different manifestations of non-radial oscillations: time-like variations and nodal lines sweeping past our sightline.  

In non-radial oscillations, a nodal line is a boundary of zero (modulated) emission that separates two regions of opposite phase.  The changing pulsation phase associated with rotating nodal structure is familiar from studies of rapidly oscillating AP stars.  For example, the rapidly oscillating AP star HR 3831 shows phase changes associated with rotation because it has a pulsation axis aligned to the magnetic axis of the star but misaligned to the rotation axis.  The right panel in Figure \ref{fig:EdwardsKurtz} shows pulsational phase changes in HR 3831 as the star's rotation carries a nodal line around the star.  Compare this to the phase behavior of the subpulses in PSR B0320+39 in the left panel.  PSR B0320+39  has a minimum in the subpulse amplitude envelope that corresponds to a $180^{\circ}$ shift in the phase, as expected for a nodal line and as seen in HR 3831.  \markcite{edw04}{Edwards} (2004) has since published more complex observed phase behaviors that are challenging both to drifting spark models and pulsation models.  We will defer the analysis of these more complex behaviors until our model has been demonstrated on a simple case, PSR B0943+10.
\clearpage
\clearpage

Because it has simple subpulse phase behavior, PSR B0943+10 is an ideal candidate for testing our model.  It has a simple average pulse shape and a single subpulse frequency.  It has been interpreted previously in the context of the drifting spark model \markcite{des01}({Deshpande} \& {Rankin} 2001), and in this paper we use the same data \markcite{sul98}({Suleymanova} {et~al.} 1998) to fit PSR B0943+10 quantitatively using our non-radial oscillation model.  As a prelude to fitting the data, we conducted an independent analysis.  In this process, we found significant behavior not mentioned in previous publications and challenging to explain with any model.  This behavior includes a splitting in the subpulse frequency and a bifurcation of the driftband in only part of the pulse profile.  Detecting these behaviors requires Fourier transforms with higher frequency resolution and driftband plots with finer amplitude resolution than previously published for this data set.  

Time series analysis of PSR B0943+10 is difficult because of a large, apparently-stochastic pulse amplitude distribution that is intrinsic to the star and not due to instrumental noise.  It is possible to filter this in Fourier space as \markcite{edw02}{Edwards} \& {Stappers} (2002) have implemented, but this may alter the data in the time domain in ways that are not intuitive.  Because most of the previous analysis of PSR B0943+10 was conducted in the time domain \markcite{des01}({Deshpande} \& {Rankin} 2001); we have chosen to do the same.  It would be a profitable exercise, but beyond the scope of this paper, to repeat the analysis using the techniques of \markcite{edw02}{Edwards} \& {Stappers} (2002).  To conduct our fitting in the time domain, we have normalized the pulses as described in \S\ref{strategy}.  The normalization process prevents our fit from being dominated by stochastic amplitude variations but also removes any useful information that might be carried by pulse amplitudes.  We are careful throughout this paper to distinguish between presentations of raw data and data that have been normalized.  In \S\ref{data}, the data analysis section, for instance, we present plots based only on data without any alteration.


Our work in this paper is organized as follows: We present a detailed summary of the model we will fit in \S\ref{model}.  We then analyze the 430 MHz data of PSR B0943+10 in \S\ref{data}, highlighting differences in our analysis to those of \markcite{des01}{Deshpande} \& {Rankin} (2001).  We then fit normalized data in the time domain to our pulsational model using Gaussfit, a robust least-squares approximation package in \S\ref{ModelFitting}.  Using Gaussfit we are able to fit quantitative values for all of the relevant parameters in our model.  We find that the parameters can be separated into geometrical and pulsational parameters and the geometrical parameters are mostly independent from the pulsational parameters. To test the quality of our fits we produce simulated data and fitted parameters and compare them to the data in \S\ref{synthetic}.  Our model successfully reproduces the essential features of PSR B0943+10.  In \S\ref{conc2}, we discuss the results of our fit and the comparison of our model to data.

\section{Our Non-Radial Oscillation Model}
\label{model}

In \markcite{cle04}{Clemens} \& {Rosen} (2004) we introduced a oblique pulsator model \markcite{kur82}({Kurtz} 1982) of high spherical degree (\el) to explain the phenomena of drifting subpulses in radio pulsars.  Originally, this model only attempted to reproduce the behavior of the total intensity (Stokes $I$) of pulsars.  In our companion paper \markcite{cle07}({Clemens} \& {Rosen} 2007), we expanded this model to encompass linear polarization by introducing two orthogonal polarization modes associated with pulsations.  The first of these orthogonal polarization modes is modulated by pulsational displacements.  These displacements have a transverse electric field vector that points toward the magnetic pole and follows the single vector model of \markcite{rad69}{Radhakrishnan} \& {Cooke} (1969).  We refer to this radiation as the ``displacement polarization mode'' and express the time dependent amplitude of this radiation as the positive portion of the function:

\begin{equation}
\label{eqn:dpm}
A_{DPM}(t) = a_{0_{DPM}} + a_{1_{DPM}} \Psi_{l,m=0}(\theta_{mag})\cos(\omega{t}-\psi_0-\psi_{delay}))
\end{equation}
\\where $\Psi_{l,m=0}$ is a spherical harmonic of high \el~ and $m=0$.  The variable $\theta_{mag}$ refers to the magnetic co-latitude, because the pulsations in our model are aligned to the magnetic pole.  The amplitudes $a_{0_{DPM}}$ and $a_{1_{DPM}}$ are to be fitted to the data.   

The pulsational displacements and their associated velocities move toward and away from the magnetic pole.  Thus, the induced electric field as a result of the velocities ($\vec{E} = \vec{v}\times\vec{B}$) is naturally orthogonal to the Radhakrishnan and Cooke vector.  We connect this induced electric field to the second orthogonal polarization mode which we refer to as the ``velocity polarization mode''. Mathematically, we express the velocity polarization mode as a time-varying amplitude:

\begin{equation}
\label{eqn:vpm}
A_{VPM}(t) = a_{0_{VPM}}{{\frac{\partial{\Psi_{l,m=0}}}{\partial{\theta_{mag}}}}}\sin(\omega{t} - \psi_0),
\end{equation}
\\which incorporates the time derivative and the $\theta_{mag}$ derivative of Equation \ref{eqn:dpm}, as appropriate for horizontal pulsation velocities.  This equation is analogous to the $V_{\theta}$ in equation three of \markcite{dzi77}{Dziembowski} (1977). 

To convert the amplitudes in Equations \ref{eqn:dpm} and \ref{eqn:vpm} into quantities that can be directly comparable to observations, we use the following transformations to calculate Stokes parameters in the frame of the star:

\begin{equation}
\label{eqn:I}
I = <A_{DPM}>^2 + <A_{VPM}>^2
\end{equation}

\begin{equation}
\label{eqn:Q}
Q' = <A_{DPM}>^2 - <A_{VPM}>^2
\end{equation} 

\begin{equation}
\label{eqn:U}
U' = 0
\end{equation}

We have followed \markcite{des01}{Deshpande} \& {Rankin} (2001) in their use of prime notation to indicate measurements in the non-rotating frame of the star.  Translating from the primed quantities into the observer's frame requires incorporating the changing longitude we observe as the star spins and imposing rotation of the polarization angle so that the polarization of the displacement polarization mode follows the changing direction of the magnetic pole, given as:

\begin{equation}
\label{eqn:chimodel}
\chi_{model} = \chi_o + \atan{\frac{\sin(\alpha)\sin(\phi-\phi_o)}{{\sin(\alpha+\beta)\cos(\alpha)-\cos(\alpha+\beta)\sin(\alpha)\cos(\phi-\phi_o)}}}
\end{equation}

Because the total linear polarization does not change with the frame of reference:

\begin{equation}
L = \sqrt{Q'^2 + U'^2} = \sqrt{Q^2 + U^2}
\end{equation}
\\then $Q'$ and $U'$ can be rotated into $Q$ and $U$ using the following transformation:

\begin{equation}
Q = L\cos(2\chi_{model})
\label{eqn:Qmodel}
\end{equation}

\begin{equation}
U = L\sin(2\chi_{model})
\label{eqn:Umodel}
\end{equation}
\\And we have to make the rotational longitude, $\phi$, and the magnetic co-latitude, $\theta_{mag}$, functions of time as follows:

\begin{equation}
\label{eqn:phi}
\phi = {\frac{t-t(\phi_o)}{P_1}} 360
\end{equation}

\begin{equation}
\label{eqn:theta}
\theta_{mag} = \cos^{-1}(\sin(\alpha)\cos(\phi-\phi_o)\sin(\alpha+\beta)+\cos(\alpha)\cos(\alpha+\beta))
\end{equation}

\subsection{\textit{The Observed Pulse Window}}
\label{window}

Our model treats the pulsations as global oscillation modes of the neutron star, which means that they can modulate emission coming from any location on the star.  However, pulsars are observed to emit radiation only from the region surrounding the magnetic poles.  The theoretical explanation for this is generally framed around the \markcite{gj69}{Goldreich} \& {Julian} (1969) aligned rotator model, in which charged particles can escape only from the polar regions.

In order to make quantitative comparisons between our model and observations, we have to impose an emission ``window'', analogous to the observed ``pulse window'', that is separate from the pulsation model and limits the effects of pulsations to the regions that are observed to emit.  For this purpose we have imposed upon $I$, $Q$, and $U$ a ``window function'' that is zero in those portions of the pulsar spin when the star is off, and is a Gaussian function with a maximum of unity in the emitting region.  Observations show that this window is not necessarily centered on the longitude of the magnetic pole \markcite{joh05}({Johnston} {et~al.} 2005). The window we use has a maximum ($\phi_{mean}$) and a width ($\sigma$) equal to the best Gaussian fit to the average pulse shape.  In all of the presentations of data that follow, zero in longitude refers to the center of this window rather than \phio, the longitude of the magnetic pole.

\section{Data Analysis of PSR B0943+10}
\label{data}

In this section we discuss our analysis of archival data of PSR B0943+10.  We highlight features in the data not mentioned in \markcite{des01}{Deshpande} \& {Rankin} (2001).  These include a small change in the subpulse frequency, a splitting in the driftband, and effects of the stochastic pulse height distribution.  Most significantly, \markcite{des01}{Deshpande} \& {Rankin} (2001) detect a modulation of the subpulses in a small portion of the data that bolsters support for their drifting spark model.  We will show in this section that the amplitude modulation, while present in that specific subsection of the data, is probably the result of the large stochastic pulse amplitude variations intrinsic to the pulsar.  These amplitude variations make the data difficult to model and in a subsequent section, \S\ref{strategy}, we will discuss our strategy for circumventing these problems. 

\clearpage
\begin{figure}
\begin{center}
\includegraphics{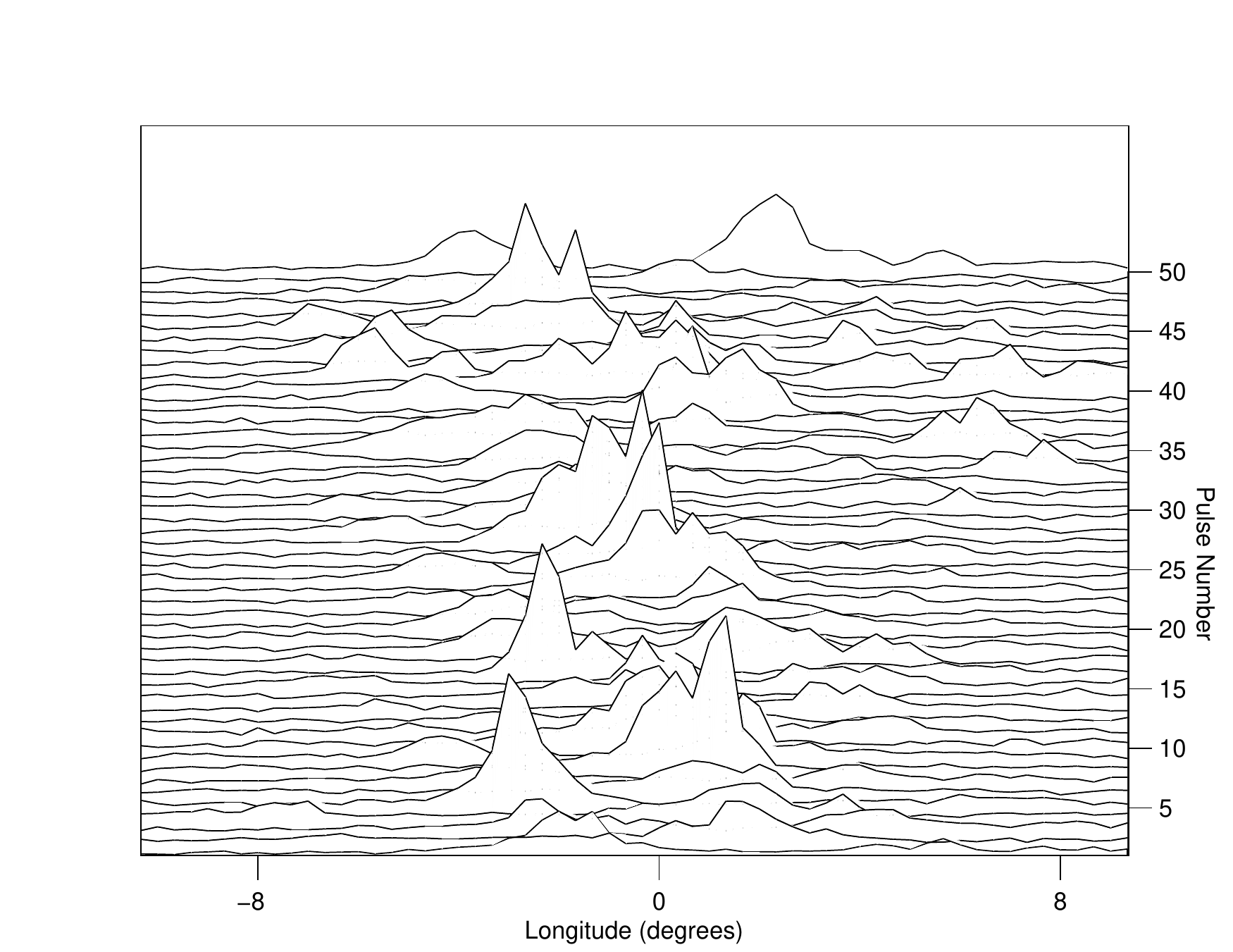}
\end{center}
\caption{Single pulses of PSR B0943+10, showing only the central $20^{\circ}$ of longitude.  The pulses are stacked at \Pone~= 1.097608 seconds, the spin period of the pulsar.  The spacing between subpulses, \Ptwo, is about 31.78 milliseconds.  \Pthree~ is identified by the Fourier transform, as calculated by \markcite{des01}{Deshpande} \& {Rankin} (2001).}
\label{fig:raw0943definitions}
\end{figure}
\clearpage

The 430 MHz data of PSR B0943+10 that we examine in this paper are the same data analyzed by \markcite{des01}{Deshpande} \& {Rankin} (2001) and include 986 pulses in all four Stokes parameters, sampled synchronously with the period of the star, \Pone~= 1.097608 seconds, in millisecond increments.  Therefore, each pulse sample contains 0.1559 seconds of measurements, corresponding to $50^\circ$ of longitude.  Of this, we designate a pulse window of about $20^\circ$, and the remaining portion of the sample is considered off-pulse, instrumental noise.  Figure \ref{fig:raw0943definitions} shows a plot of the central $20^{\circ}$ of single pulses of PSR B0943+10.  The pulses clearly drift through the pulse window.  Each pulse sample is not contiguous with the adjacent pulses, and there are not an integer number of time intervals, $\delta t$, between each pulse.  Our analyses and that of \markcite{des01}{Deshpande} \& {Rankin} (2001) assume the signal from the pulsar is zero during the offpulse portion, but this is not shown in Figure \ref{fig:raw0943definitions}.  Figure \ref{fig:raw0943definitions} illustrates the important periods in pulsar terminology: \Pone, the spin period of the pulsar, the time between pulses, and the period at which the data are folded in Figure \ref{fig:raw0943definitions}; and \Ptwo, the spacing between adjacent subpulses.  \Pthree~ is ordinarily the length of time for a subpulse to return to a given longitude, but we have followed \markcite{des01}{Deshpande} \& {Rankin} (2001) and used \Pthree~ from the lowest alias of \Ptwo~ in the Fourier transform, specifically \Pthree~ = 1.867\Pone.

The data were originally acquired by \markcite{sul98}{Suleymanova} {et~al.} (1998) who describe the observing setup.  The subpulses appear in their steady, highly organized bright ``B''-mode state for the first 816 pulses, where the spacing between the subpulses, \Ptwo, is about 31.78 milliseconds.  The remaining 170 pulses are disorganized with generally lower amplitude and the pulsar is considered to be in its ``Q''-mode (``quiescent'') state.  We do not present a full repetition of the analysis of \markcite{des01}{Deshpande} \& {Rankin} (2001), nor do we apply all of their analysis techniques, e.g. folding the data at 20\Pthree, where \Pthree~ (1.866\Pone) is the length of time for a subpulse to return to a given longitude.  Instead, we refer to the reader to their paper for most of the data analysis, except for those instances where we find significant behavior of the star not shown in their analysis.  

\markcite{des01}{Deshpande} \& {Rankin} (2001) use four basic tools in their analysis: synchronous folding of the data, the Fourier transform, the longitude resolved fluctuation spectrum, and the harmonic resolved fluctuation spectrum.  When synchronously folding the data at a given period, usually \Pthree, the subpulses drift through the pulse window at each successive spin until they repeat in longitude, thus creating a driftband plot.  For example, \markcite{des01}{Deshpande} \& {Rankin} (2001) fold the data at 1.867\Pone~(=\Pthree) and at 37.346\Pone~(=20\Pthree), in their figures 8 and 9, respectively.  The longitude resolved fluctuation spectrum, first introduced by \markcite{bac73}{Backer} (1973), is a Fourier transform calculated at each longitude.  The harmonic resolved fluctuation spectrum is a Fourier transform of the time series stacked at the spin frequency.  \markcite{edw02}{Edwards} \& {Stappers} (2002) have shown that the combination of the longitude and harmonic resolved fluctuation spectra present the same information as a two-dimensional Fourier spectrum.  While we have also calculated the two-dimensional Fourier spectrum, we do not present it here because it is less familiar and not directly comparable to the analysis of \markcite{des01}{Deshpande} \& {Rankin} (2001).

\clearpage
\begin{figure}
\begin{center}
\includegraphics{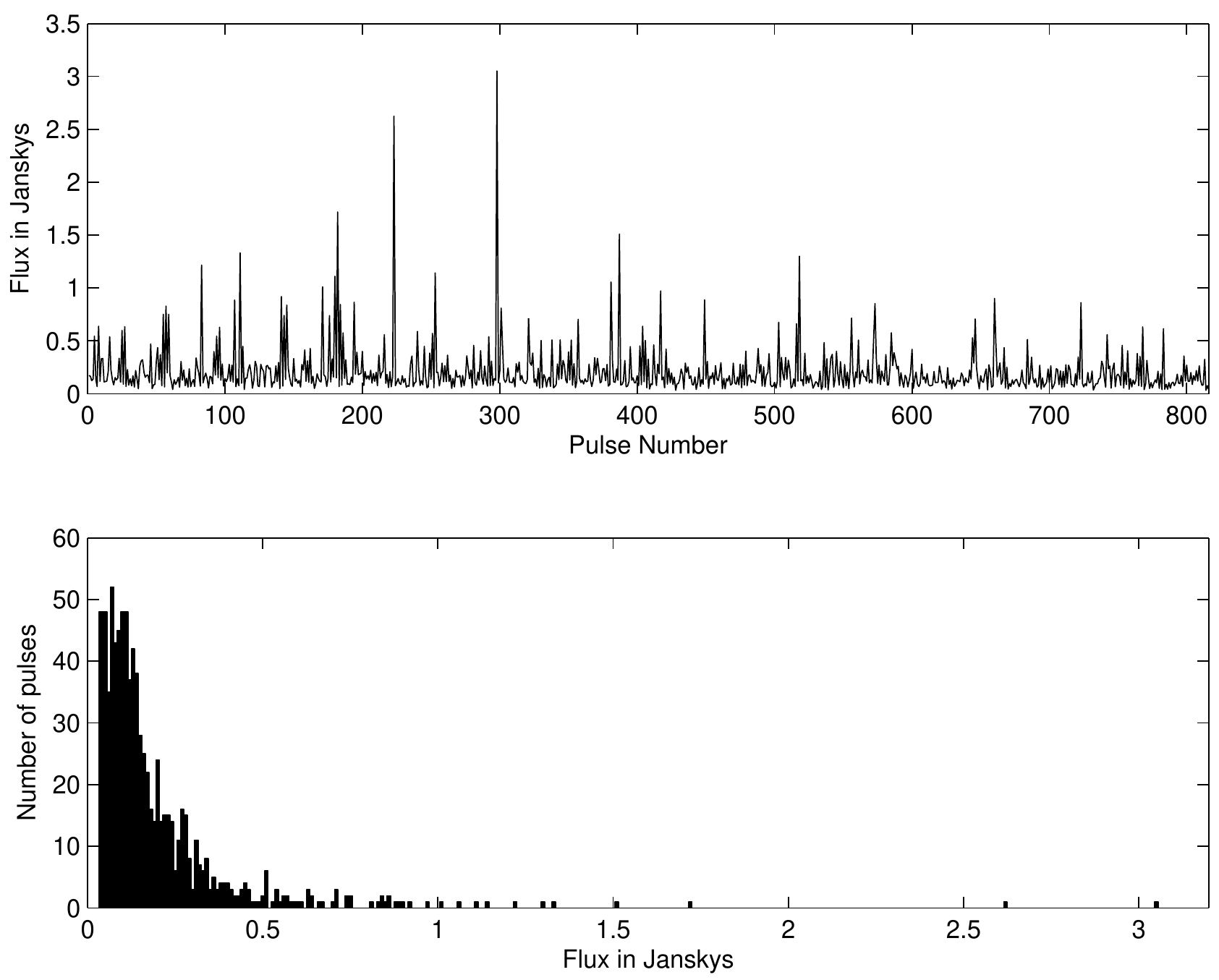}
\end{center}
\caption{The light curve of PSR B0943+10 (top panel) showing large pulse-to-pulse variations in amplitude.  The lower panel shows the pulse height distribution for the data in the top panel.}
\label{fig:amplitudes}
\end{figure}
\clearpage

The most difficult aspect of modeling PSR B0943+10 is the large stochastic variation in the pulse amplitudes.  The average signal-to-noise of an integrated pulse is 4.93, where we used the off-pulse instrumental noise as the noise estimate.  The standard deviation of the onpulse data average to be 7.13 of the standard deviation of the noise, but varies from 1.23 to 109.87.  From this, we conclude the large variations in pulse height, as shown by the pulse height distribution given in Figure \ref{fig:amplitudes}, are due to stochastic variations of the pulsar, not instrumental noise.  Indeed, \markcite{cor78}{Cordes} (1978) has argued that pulsar subpulses are amplitude modulation of shot noise-like emission.  Our analysis of the data in this section is tempered by the possible effects of the stochastic variation, but we do not attempt to remove this variation by any normalization process until \S\ref{ModelFitting}.


\subsection{\textit{Driftbands}}
\label{driftbands}

We begin our analysis by folding the data at \Pthree, a technique used by \markcite{des01}{Deshpande} \& {Rankin} (2001) to show changes in the shape of individual subpulses as a function of longitude.  \markcite{des01}{Deshpande} \& {Rankin} (2001) remove an ``aperiodically fluctuating base'' prior to presenting their data.  They do not describe their method for removing this base in enough detail for us to reproduce it.  We have decided to present the unaltered data which may not be exactly the same as in their presentation.

\begin{figure}
\begin{center}
\includegraphics{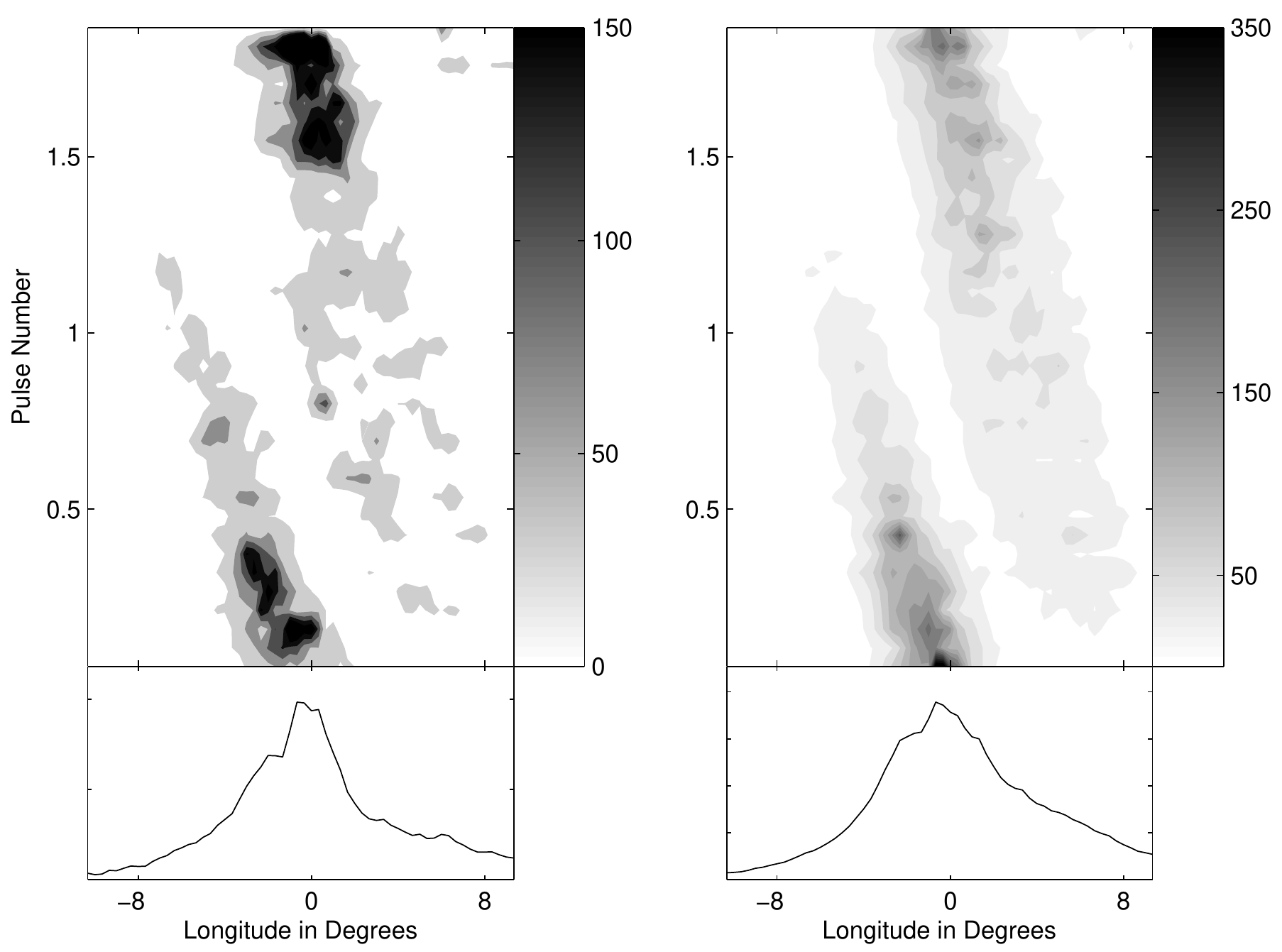}
\end{center}
\caption{Left panel: The first 100 pulses of PSR B0943+10 folded at \Pthree~ = 1.86598\Pone~ seconds.  The value of \Pthree~ was calculated based on the value of \Ptwo~ = 31.7816 milliseconds, taken from the Fourier transform of the entire 816 pulses, and using equation 4 from \markcite{cle04}{Clemens} \& {Rosen} (2004) (${{1}\over{P_3}} = {{{1}\over{P_{time}}} - {{n}\over{P_1}}}$, where $nP_{time} \approx P_1$).  Right panel: All 816 pulses folded at the same value of \Pthree. The amplitude of the greyscale is in mJy.}
\label{fig:driftdata}
\end{figure}

Our driftband plots in Figure \ref{fig:driftdata} show broadening, or even splitting, on the right-hand-side of the profile that is not evident in the lower resolution plot in \markcite{des01}{Deshpande} \& {Rankin} (2001).  To explore this, we folded shorter 100-pulse segments of the data at \Pthree, and noticed an obvious splitting on the right side of driftband as shown in the left panel of Figure \ref{fig:driftdata}.  While we only show the first 100 pulses here, driftbands of other 100-pulse segments of the data also display splittings on the right-hand-side of the driftband.  The washed out appearance of the split in the driftband in the full data set (right panel in Figure \ref{fig:driftdata}) is a result of the drifting in the subpulse frequency, which we will discuss in \S\ref{qsft}.
\clearpage
\clearpage

The splitting in the driftband is difficult to explain because it is a longitude-dependent phenomenon.  In the model of \markcite{des01}{Deshpande} \& {Rankin} (2001) in particular, there is a pattern of 20 sub-beams circling the magnetic pole.  In each successive spin of the star, an individual sub-beam moves a few degrees earlier in longitude.  This means the same sub-beam as seen on the right side of the pulse window will, in the next spin of the star, have moved to the left side.  The splitting in the driftband indicates that all sub-beams must be split when they are present on the right side of the pulse window and that they recombine when they move to the left side.  This behavior is difficult to explain with the rotating spark model without the multiplication of new quantities, but we will show it is a natural result of the phase difference between the displacement and velocity polarization modes in our model.

\subsection{\textit{Fourier Transform}}
\label{qsft}

The Fourier transform is an indispensable tool for finding frequencies in noisy data.  \markcite{des01}{Deshpande} \& {Rankin} (2001) use a Fast Fourier Transform to determine the subpulse frequency, but they do not calculate a fully-resolved transform of all 816 pulses analyzed in their paper. Instead they calculate Fast Fourier Transforms of 256-pulse sections and average them.  This prevents them from seeing fine structure in the subpulse frequency that is only visible at high resolution.  We will show that our resolved Fourier transform of the entire 816 pulses shows splitting in the subpulse frequency.  This splitting is not the same as the sidelobes uncovered by \markcite{des01}{Deshpande} \& {Rankin} (2001); this spacing is significantly smaller and does not have the obvious symmetry to be sidelobes due to an amplitude modulation.  Our analysis will show that it is impossible with the limited data to determine whether the two largest peaks are two closely-spaced, independent frequencies or the result of instabilities of the subpulse frequency.  The presence of two independent frequencies would be a significant complication for the drifting spark model but would convey asteroseismological information if the variations are non-radial oscillations.  Additional data is necessary for a definitive identification of frequency splitting.
\clearpage
\begin{figure}
\begin{center}
\includegraphics{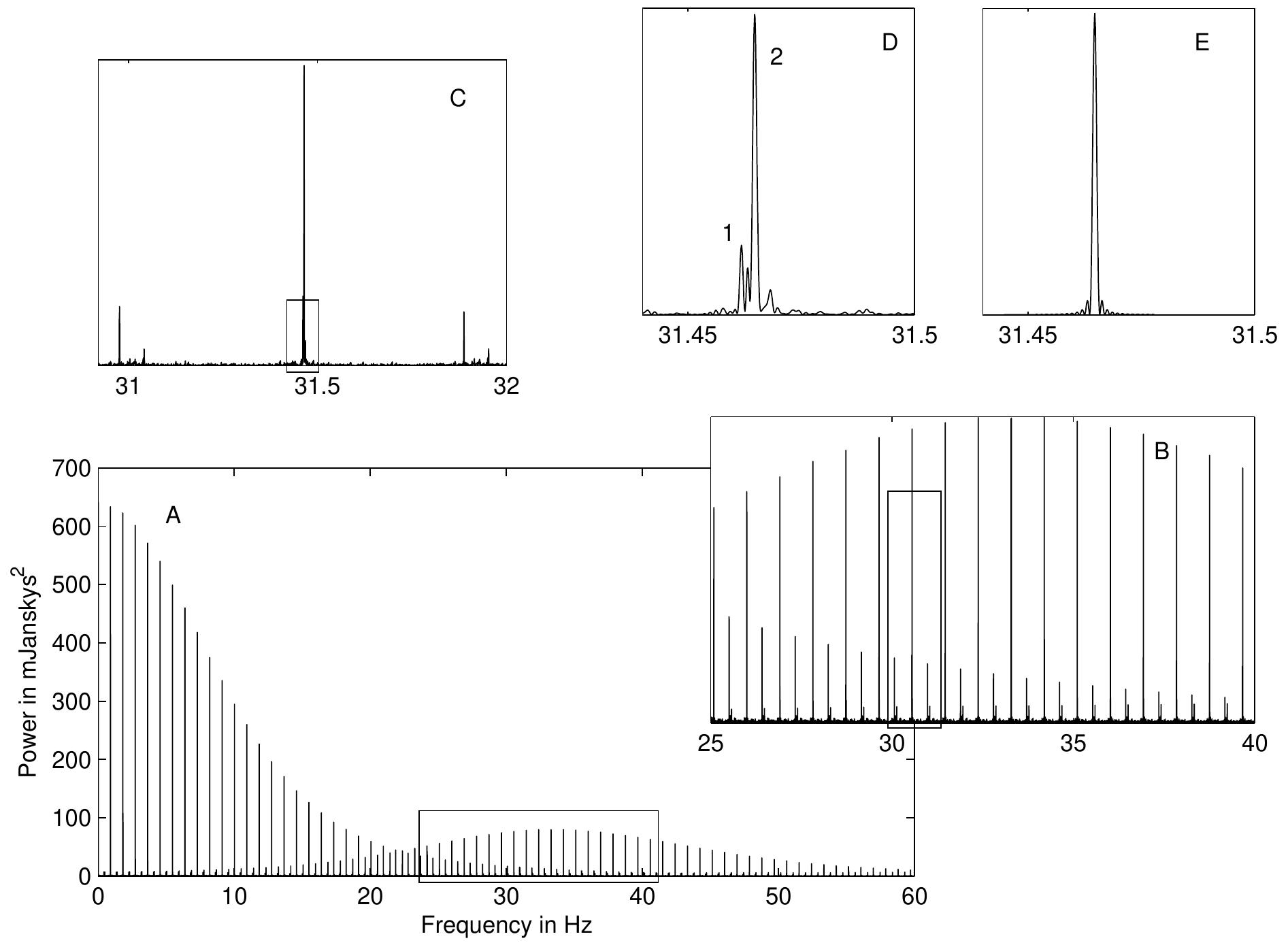}
\end{center}
\caption{A fully-resolved Fourier transform of all 816 pulses.  Panel (A) shows the transform up through the first 60 harmonics.  Each successive panel enlarges the area of the transform, eventually focusing on the subpulse frequency in panel (D).  The subpulse frequency is split into several peaks; the largest two are labeled.  The spacing between the two labeled peaks is about three times greater than the resolution of the Fourier transform.  For reference, panel (E) is the Fourier transform on a synthetic single-frequency sine wave sampled the same as the data.}
\label{fig:FreqSplitting}
\end{figure}
\clearpage 
To compute our Fourier transforms we use a tool written by Carl Hansen for the Whole Earth Telescope \markcite{nat90}({Nather} {et~al.} 1990).  It is optimized for discrete, contiguous chunks of data with gaps in between.  Unlike the Fast Fourier Transform, this tool does not require an integer number of time intervals, $\delta{t}$, between the chunks and is therefore appropriate for the data from PSR B0943+10.  Like the Fast Fourier Transform, this tool assumes zeros between chunks of data, but as a discrete Fourier transform, it does not require $2^n$ datapoints or a resampling the data into equally spaced time bins.  The bottom panel (A) of Figure \ref{fig:FreqSplitting} shows the Fourier transform up through the 60th harmonic of all 816 pulses of PSR B0943+10, increasing the resolution by a factor of approximately three over a Fourier transform using only 256 pulses.  A closer inspection of the subpulse frequency in the top panel (D) of Figure \ref{fig:FreqSplitting} shows that the peak is split into several peaks, and the largest four are all statistically significant.  For reference, panel (E) of Figure \ref{fig:FreqSplitting} shows a window function which is the Fourier transform of a synthetic single-frequency sine wave sampled as the data.  

The peaks in panel (D) are representative of all the aliased subpulse peaks.  The larger of the two largest peaks (peak 2) is at 31.4647 Hz (\Ptwo~= 31.7816 milliseconds), which is not the largest subpulse alias in panel (A).  We have chosen this subpulse alias as the subpulse frequency because this alias shows an integer relation to the second and third harmonics which are not shown in Figure \ref{fig:FreqSplitting}.   The location of the smaller peak (peak 1) is at 31.4618 Hz (\Ptwo~= 31.7846 milliseconds); the difference in the two peaks is 0.0029 Hz.  This difference in frequency would be barely visible in a 256-pulse Fourier transform, especially if several transforms were averaged together which could broaden the subpulse frequency.

To explore frequency and phase stability of the subpulses, we divided the data into sequential 50-pulse segments, where the number of pulses in each segment (50) was chosen arbitrarily.  Since we were fitting only the subpulses in the pulse window, we needed enough pulses that the subpulse frequency could be accurately fit.  The upper limit on our segment length was determined by attempting to minimize the amount of the frequency wander and having enough data segments to determine any trends. 

Fourier transforms of these smaller subsets of data can still clearly show the subpulse frequency, but they do not have the frequency resolution to separate the subpulse into the distinct peaks seen in the fully-resolved Fourier transform.  We used Gaussfit \footnote{http://clyde.as.utexas.edu/Software.html}, a robust least-squares approximation package, to fit the intensities (Stokes $I$) in each of the of 50-pulse segments to a periodic function.  Because the variance in the data is much higher than the off-pulse instrumental noise, we use the variance as an error estimate.  This artificially forces the reduced $\chi^2$ to be near one, so we have no absolute measure of the goodness of our fits.

\begin{figure}
\begin{center}
\includegraphics{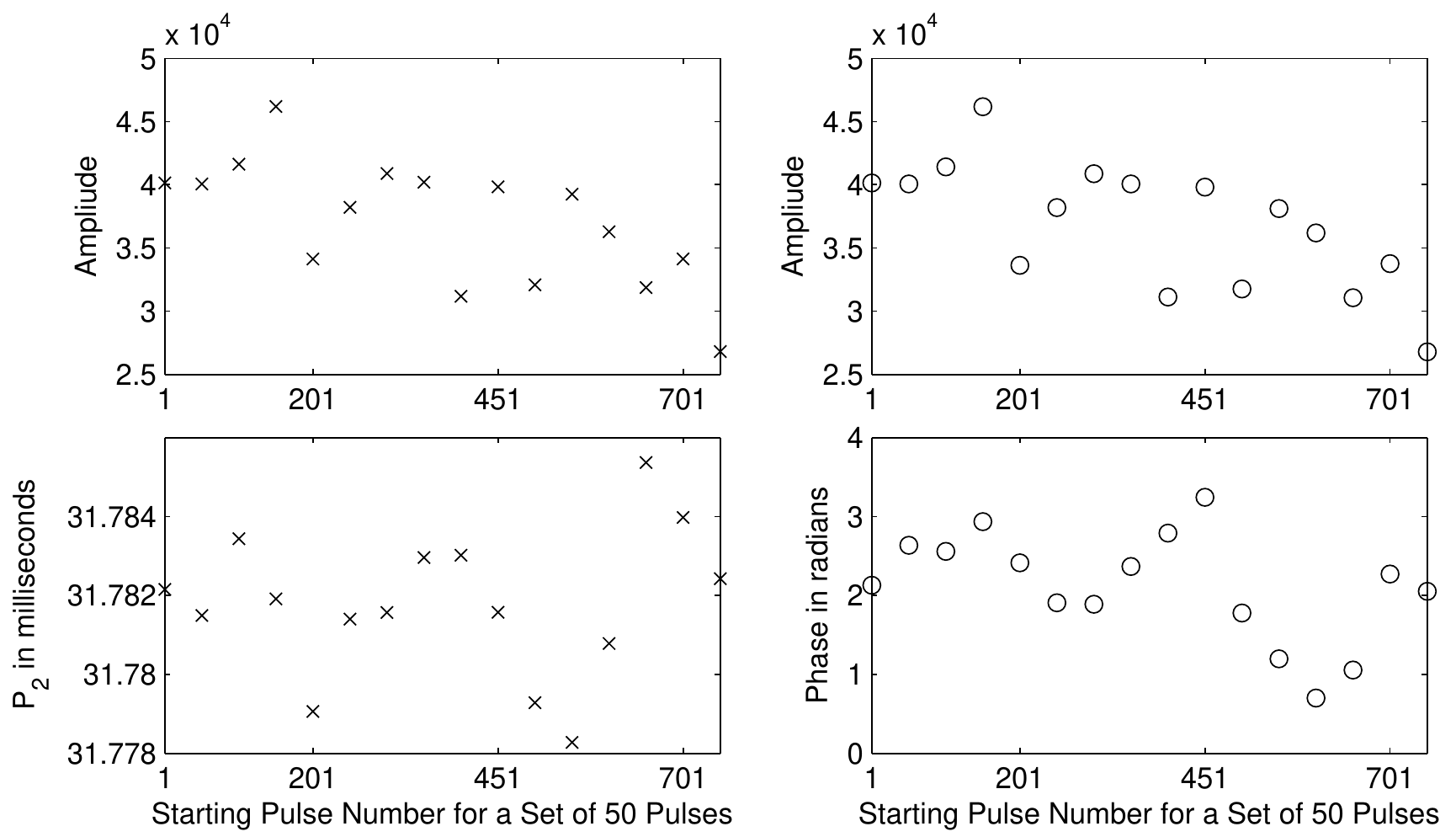}
\end{center}
\caption{Left panels: The results of fitting consecutive 50-pulse segments of the data with the positive portion of the function $I = A\cos(\frac{2\pi{t}}{P_2}-\psi)$.  Top left panel: The fitted value of the amplitude for each 50-pulse segment.  Bottom left panel: The corresponding fitted period for each 50-pulse segment.  Right panels: The results of fitting a single sine wave with fixed subpulse period, \Ptwo, to consecutive 50-pulse segments of the data.  Top right panel: The fitted value of the amplitude for each 50-pulse segments using Gaussfit.  Bottom right panel: The corresponding fitted phase for each 50-pulse segments in radians.  The error bars are not included because the errors are sufficiently small that they are not visible within the resolution of the plot.}
\label{fig:GaussfitRunningFT}
\end{figure}

We cannot use a sine wave alone for this fit because its mean would be zero and our data are all positive.  Removing the mean from the data, which would be the normal procedure for studying pulsations, does not help in this case because of the large amplitude variations.  Thus we have used for our fit the positive portion of our function $I = A\cos(\frac{2\pi{t}}{P_2}-\psi)$.  This yields good convergence from Gaussfit and permits us to measure the period and phase in each segment of the data.  We have plotted the period and amplitude as a function of data segment in the left panels of Figure \ref{fig:GaussfitRunningFT}. 

In \markcite{cle04}{Clemens} \& {Rosen} (2004), we pointed out that \Ptwo~ originally defined as the time between adjacent subpulses in a driftband, may not be a good estimate of the underlying pure subpulse frequency in our pulsation models.  Consequently we called subpulse period \Ptime.  \markcite{edw06}{Edwards} (2006) has pointed out that the amplitude windowing which distorts measurements of \Ptwo~ in individual pulses does not affect measurements that rely on subpulse phase.  In this paper, all of our measurements of \Ptwo~ are either from Fourier transforms or linear fits using pure sine waves, which means our estimate of \Ptwo~ is a good measurement of \Ptime~ and we are using the more familiar notation of \Ptwo~ for the measured subpulse period.

Our analysis of the smaller data segments reveals that at low resolution, \Ptwo~ wanders over the entire run.  The change in \Ptwo~ with subset is shown in the bottom left panel of Figure \ref{fig:GaussfitRunningFT} and the corresponding amplitudes ($A$) are shown in the top left panel.  As shown in Figure \ref{fig:FreqSplitting}, the fully-resolved Fourier transform shows several distinct peaks.  Each subset in Figure \ref{fig:GaussfitRunningFT} is not long enough to resolve these peaks individually, instead they appear to be a single peak with its maximum determined by the relative sizes of the unresolved peaks, causing the wander in \Ptwo.  This single peak is large compared to the noise and therefore Gaussfit is able to fit it very accurately, resulting in the extremely small error bars in Figure \ref{fig:GaussfitRunningFT}.  The width of the peak, however, is determined by the length of the time series and is larger than the error of the fit.

The pattern in the period change is more complex than we would expect from a pure frequency splitting, which would generate a regular pattern of period changes.  We can explore things further by examining the subpulse phase.  If the subpulse frequency were solely amplitude modulated we would expect symmetric sidelobes in the Fourier transform and our least squares fit using a fixed frequency would show a stable subpulse phase.  If the subpulse is actually two closely-spaced frequencies, a fit with a fixed frequency would result in a periodic change in the subpulse phase.   
\clearpage
\clearpage

To examine the phase behavior, we averaged the subpulse frequencies in the bottom left panel of Figure \ref{fig:GaussfitRunningFT} and then fit each of the 50-pulse subsets with Gaussfit with the same positive portion of the function $I = A\cos(\frac{2\pi{t}}{P_2}-\psi)$, but with fixed \Ptwo~ = 31.7818 milliseconds.  The resulting phase as a function of time is shown in the bottom right panel of Figure \ref{fig:GaussfitRunningFT} and the corresponding amplitudes are in the top panel.  Because it appears that the subpulse frequency and phase wander through only one or two cycles, we would require a longer data stream to conclusively determine whether more than one frequency is present.  The amplitudes of the phase variations appears to increase with pulse number.  Our data terminates with a transition from the stable ``B''-mode to the disordered ``Q''-mode, where the fits in the section were confined to the 816 pulses in the ``B''-mode.  Whether the transition into a disordered state is related to the increase in size of subpulse phase and period variations is a also question requiring more data. 


\subsection{\textit{Longitude and Harmonic Resolved Fluctuation Spectra}}

Stochastic variation in the pulse amplitudes complicate attempts to extract any underlying signal for comparison to our model.  While stochastic variation averages away in the driftbands and Fourier transforms of large segments of the data, in short sections the probability is greater that large stochastic variations can mimic an amplitude modulation.  Within this data set, \markcite{des01}{Deshpande} \& {Rankin} (2001) have detected a periodic amplitude modulation in the region of pulses 129-384 of the data.  This periodic amplitude modulation manifests itself as sidelobes around the lowest alias of the subpulse frequency in the longitude resolved fluctuations spectra; these sidelobes are distinctly different that the peaks discussed in \S\ref{qsft} and shown in Figure \ref{fig:GaussfitRunningFT}.   The amplitude modulation is a key component for increasing their confidence in the circulating spark model; \markcite{des01}{Deshpande} \& {Rankin} (2001) interpret it as a persistent pattern of sub-beam brightness.  They view it as confirming the existence of twenty distinct sub-beams.  A single-mode pulsational model like ours cannot reproduce either the stochastic variations in the pulse height or the amplitude modulation that produces the sidelobes.  However, it is legitimate to ask whether the appearance of symmetric sidelobes in a subset of the data is not consistent with stochastic amplitude variations alone, rather than indicative of a periodicity in the star that should be fitted by models.

\begin{figure}
\begin{center}
\includegraphics{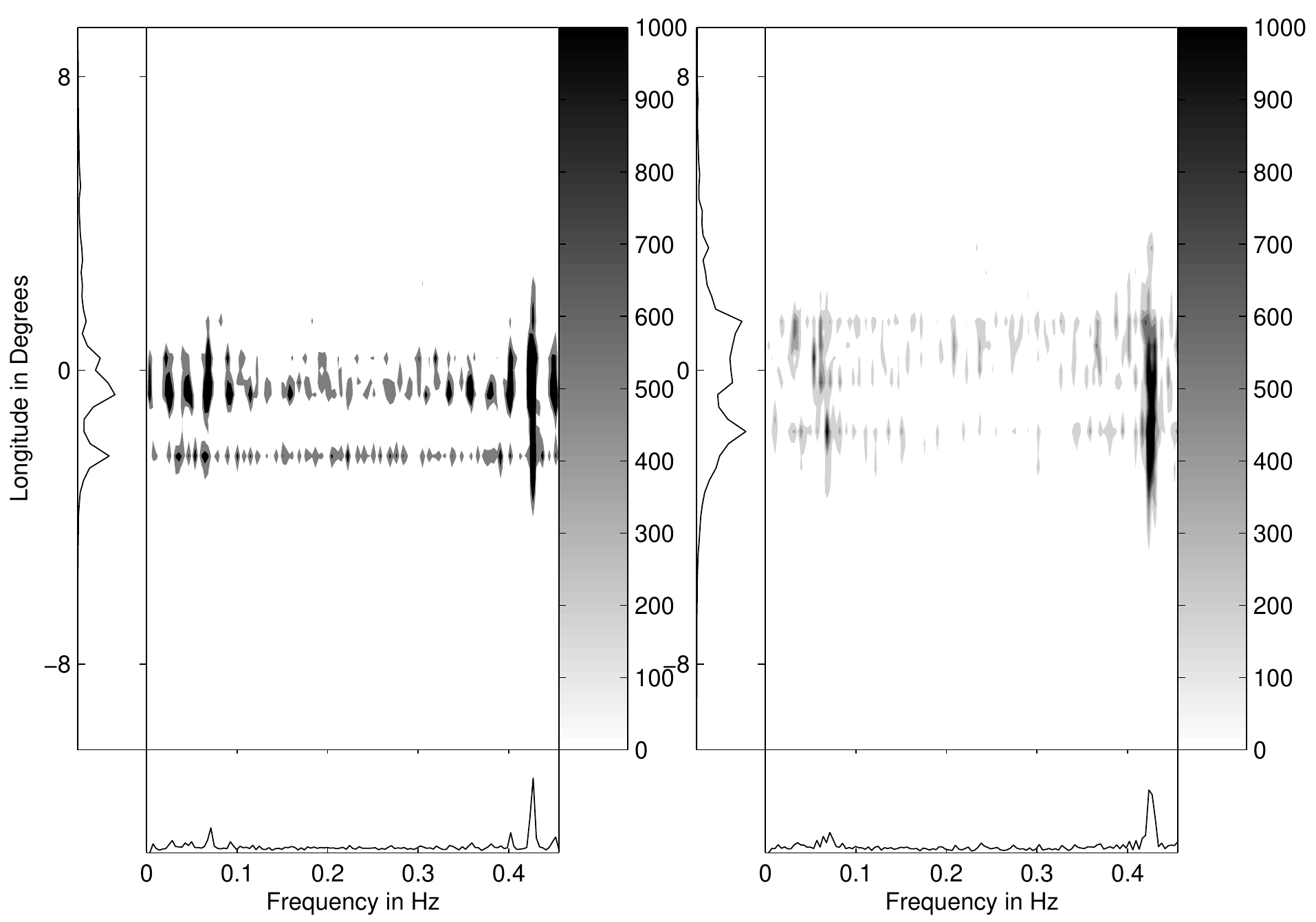}
\end{center}
\caption{The longitude resolved fluctuation spectrum for two consecutive 256-pulse segments of the data from PSR B0943+10.  Left panel: The subsection of the data (pulses 129-384) where \markcite{des01}{Deshpande} \& {Rankin} (2001) discovered the tertiary modulation that supports the rotating spark model.  Right panel: The longitude resolved fluctuation spectrum for the next portion of the data (pulses 385-640); the tertiary modulation is not present.  The contour levels are in mJy$^2$.}
\label{fig:lrf}
\end{figure}

\begin{figure}
\begin{center}
\includegraphics{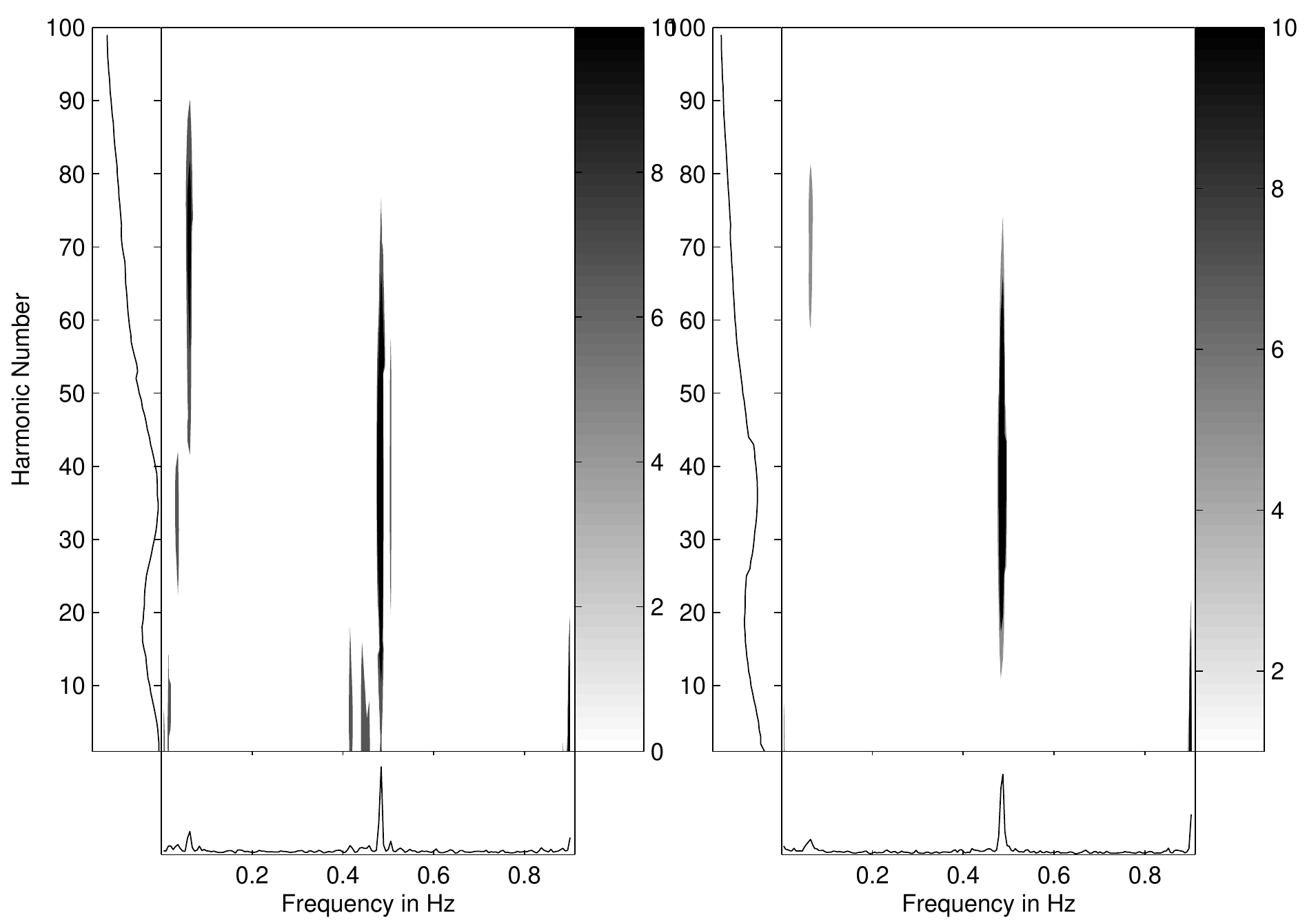}
\end{center}
\caption{The harmonic resolved fluctuation spectrum for the same two consecutive 256-pulse segments of the data from PSR B0943+10 as in Figure \ref{fig:lrf}.  Left panel: The subsection of the data (pulses 129-384) where \markcite{des01}{Deshpande} \& {Rankin} (2001) discovered the tertiary modulation.  Right panel: The harmonic resolved fluctuation spectrum for the next portion of the data (pulses 385-640).  The contour levels are in mJy$^2$.}
\label{fig:hrf}
\end{figure}

The amplitude modulation interpreted by \markcite{des01}{Deshpande} \& {Rankin} (2001) can be seen in the longitude and harmonic resolved fluctuation spectra of pulses 129-384 in the left panels of Figures \ref{fig:lrf} and \ref{fig:hrf}.  In Figure \ref{fig:lrf}, the aliased subpulse frequency is reflected about the Nyquist frequency and appears at about 0.43 Hz.  As the panel on the left shows, the subpulse signal has sidelobes indicative of amplitude modulation with a period of just over 37\Pone.  The difference in our longitude resolved fluctuation spectrum (left panel of Figure \ref{fig:lrf} and that of \markcite{des01}{Deshpande} \& {Rankin} (2001) (their figure 7) is likely due to their removal of the aperiodically fluctuating base, which we do not remove.  In Figure \ref{fig:hrf}, the aliased subpulse frequency is at its true value of 0.49 Hz and shows no significant sidelobe structure.  Visual inspection of the harmonic resolved fluctuation spectrum in figure 4 of \markcite{des01}{Deshpande} \& {Rankin} (2001) does not show evidence of sidelobes, nor do we see them in our harmonic resolved fluctuation spectrum of the entire 816 pulses.  The sidelobes are also not present in the other sections of the data, neither in the longitude or harmonic resolved fluctuation spectrum.  For comparison, we show the next 256 pulses (pulses 384-640) in the right panels of Figures \ref{fig:lrf} and \ref{fig:hrf}.
\clearpage

\clearpage

In favor of the significance of amplitude modulated sidelobes, Monte Carlo simulations in which we shuffled the amplitudes in pulses 129-384 of the data showed a very low probability (less that 1\%) that two peaks of the size seen in the longitude resolved fluctuation spectra would occur by chance.  However, the probability that an amplitude modulation could appear in the longitude resolved fluctuation spectrum and be masked by noise in the harmonic resolved fluctuation spectrum is equally small.  The sidelobes do not appear in the harmonic resolved fluctuation spectrum neither in our plots nor in the plots of \markcite{des01}{Deshpande} \& {Rankin} (2001).  There are a number of other reasons to suspect that it is something other than a pure amplitude modulation.  Based on the models of the drifting carousel pattern of sparks, \markcite{edw02}{Edwards} \& {Stappers} (2002) show that an amplitude modulation caused by a persistent patterns in the subpulses ought to appear as a sidelobe of the DC component at zero frequency.  In the same section of the data, it does not appear as a low frequency sidelobe.  It is also not present in any of the other independent 256-pulse samples nor has it reappeared in subsequent data. 

Clearly it is of the highest importance to establish whether periodic amplitude modulation occurs in the drifting subpulses.  It would be the strongest evidence in favor of the drifting subpulse model and against the model we present here. Its appearance in a single section of the data, like the appearance of frequency splitting and phase wandering in the subpulses, needs to be addressed with additional data before we can assess its consequences for our model.

\section{Model Fitting and Simulations}
\label{ModelFitting}

\begin{figure}
\begin{center}
\includegraphics{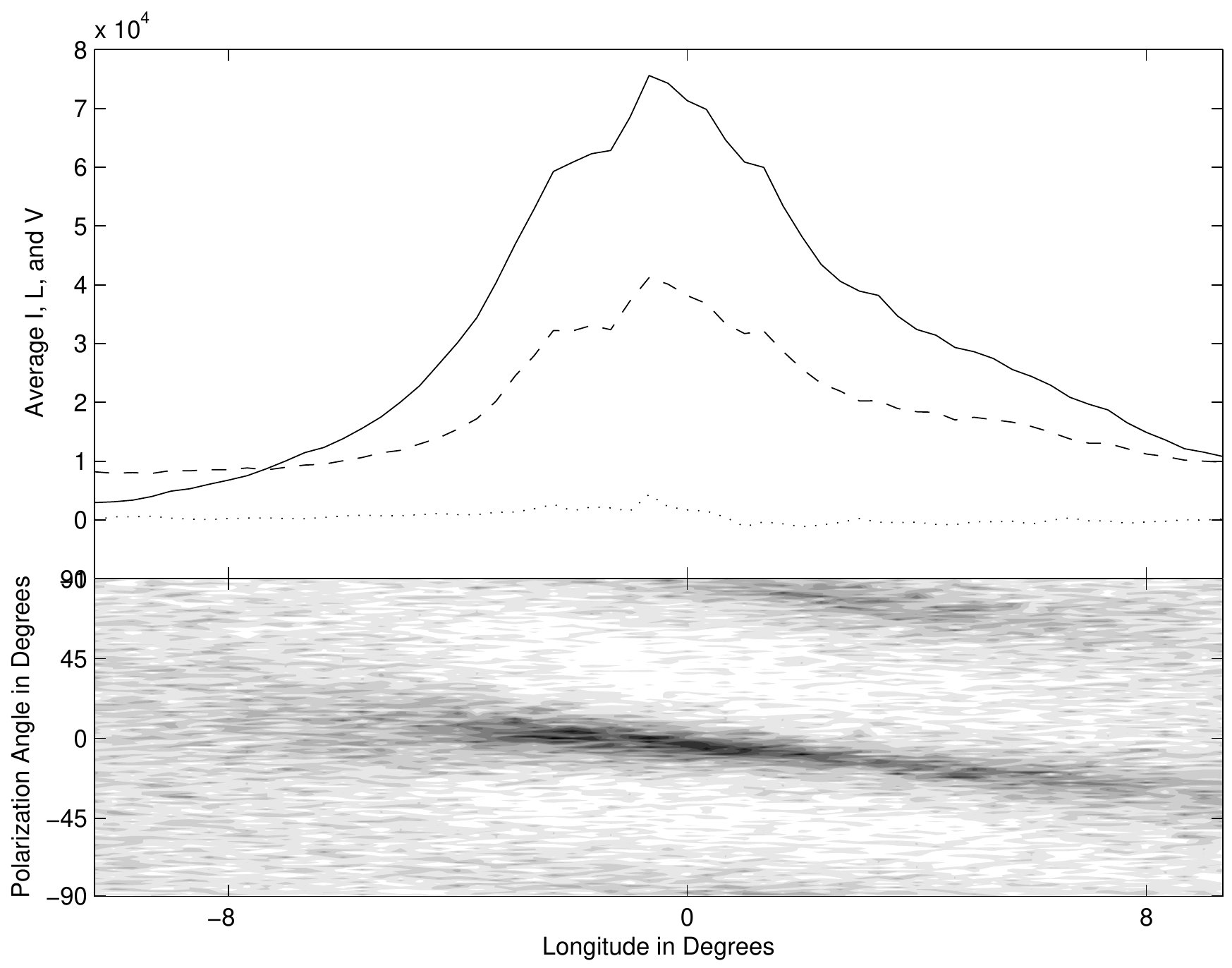}
\end{center}
\caption{Top panel: Average intensity (solid line), linear polarization (dashed line), and circular polarization (dotted line) for PSR B0943+10 at 430 MHz.  Bottom panel: Polarization angle histogram of the same data.}
\label{fig:PulseProfile}
\end{figure}

Departing now from new phenomena uncovered by our analysis, we address the central question of this work, which is whether or not our pulsational model can reproduce the stable features of the observations.  We will address this question by first attempting to fit the data using our pulsational model.  The purpose of the fitting is to select objective quantities for the model variables. Once we have completed the fitting, we will use the fitted parameters to generate synthetic pulsar lightcurves and see how well they reproduce the features of the drifting subpulses.  As we will see, the best fits to our model generate synthetic data that reproduces the driftbands with splitting, the polarization properties shown in Figure \ref{fig:PulseProfile}, and the essential features of the harmonic and longitude-resolved fluctuation spectra.
\clearpage
\clearpage
\subsection{\textit{Fitting Strategy}}
\label{strategy}

Before we can apply our general model to a specific star like PSR B0943+10, we must first determine whether orthogonal polarization mode emission is present in the star and whether the emission polarization angles are consistent with the single vector model of \markcite{rad69}{Radhakrishnan} \& {Cooke} (1969).  Figure \ref{fig:PulseProfile} shows a histogram of the distribution of polarization angles in individual subpulses as first introduced by \markcite{sti84a,sti84b}{Stinebring} {et~al.} (1984a, 1984b).  At each longitude it counts how many individual pulses show polarization angles in each bin.  The polarization angles clearly group around two parallel polarization tracks.  These tracks are only gently curved indicating that our sightline misses the magnetic pole so that the entire pulse samples a small range of magnetic longitude.  

We have identified the stronger track with the displacement polarization mode of our model and the weaker track with the velocity polarization mode because the reverse identification did not yield acceptable fits.  Inspection of the lower panel of Figure \ref{fig:PulseProfile} suggests that the amplitude coefficient for the displacement polarization mode must be larger than the velocity polarization mode, i.e. the displacement polarization mode is the dominant mode in this star.  By convention, we would call this the primary polarization mode but we want to avoid attaching the displacement polarization mode to the primary polarization because it is not clear that the displacement polarization mode will dominate in every pulsar.

\begin{figure}
\begin{center}
\includegraphics{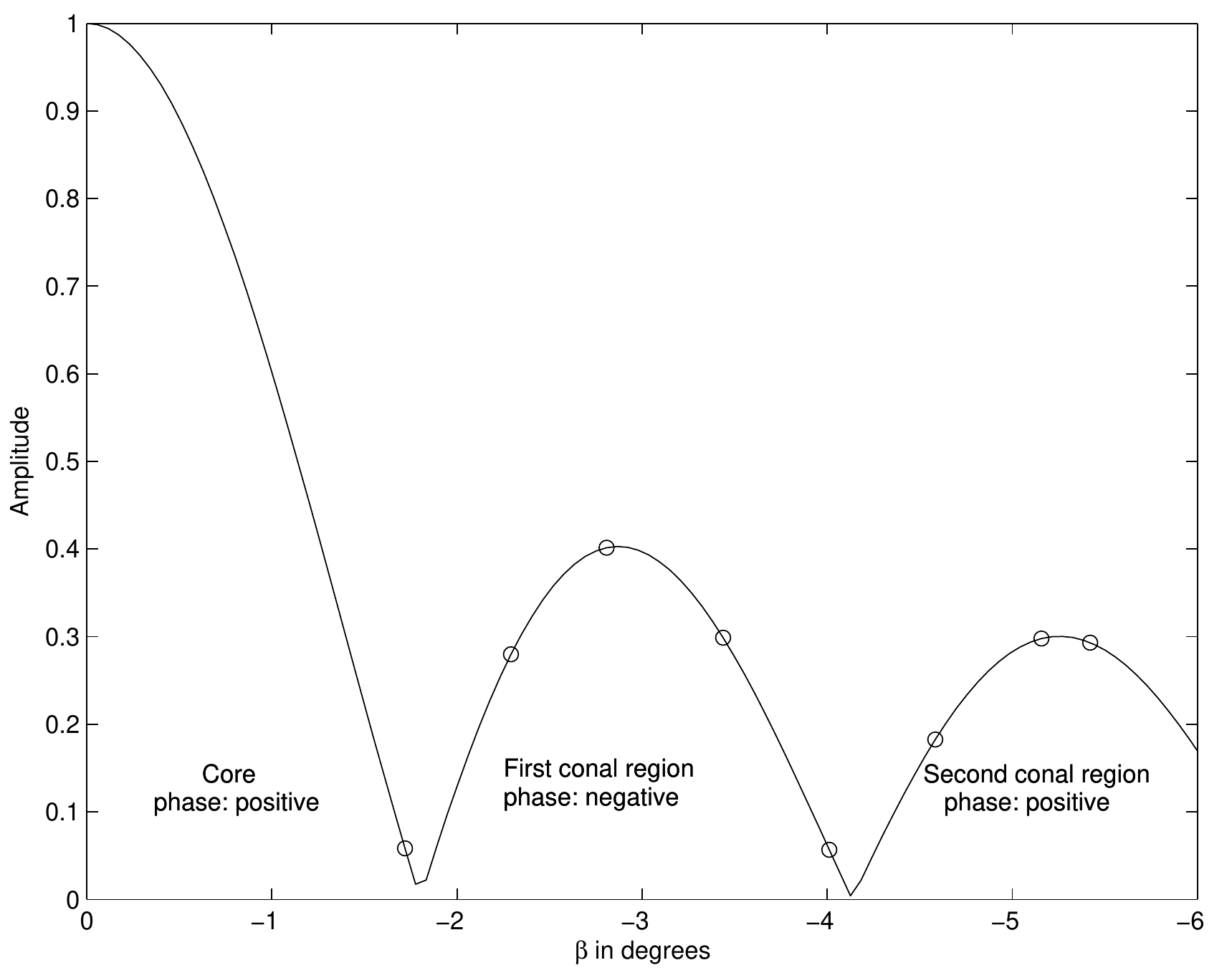}
\end{center}
\caption{A scan along a single magnetic longitude and shows the location of nodal lines for a particular \el~ (\el=75) for various choices of the impact parameter \bt.   The magnetic pole is at $\beta=0^{\circ}$, the nodes are at about $-1.8^{\circ}$ and $-4.1^{\circ}$, and the antinodes are at $-2.9^{\circ}$ and $-5.2^{\circ}$. The circles indicate values of \bt~ explored in our fits to the pulsation model.}
\label{fig:beta}
\end{figure}

We can fit the rotation axis inclination, \al, and impact parameter, \bt, using these polarization angle swings but the fits are ambiguous, yielding only the ratio of \al~ to \bt. We will discuss our results for these parameters in \S\ref{FitstoPA}.  \markcite{des01}{Deshpande} \& {Rankin} (2001) apply external information to constrain \al~ independently and for consistency we have chosen values of \al~ near theirs.  The only pulsational parameter that interacts with these (\al, \bt) is the spherical harmonic degree \el~ because the positions of the nodal lines that encircle the magnetic pole are related \el, and the path our sightline threads through these nodal lines is related to \bt.  Figure \ref{fig:beta} is a scan of the spherical harmonic along a single magnetic longitude and shows the location of nodal lines for a particular \el~ (\el=75) and various choices of the impact parameter \bt.  Using only the data on PSR B0943+10 analyzed in this paper, there is no way for us to know which nodal region our sightline passes through nor is there anyway to constrain \el~ that is independent of \bt.  In most emission models, the divergence of the dipole field magnifies the structure at the stellar surface so the apparent \el~ we fit represents a surface \el~ approximately seven times higher, as we explained in \markcite{cle04}{Clemens} \& {Rosen} (2004).  Without a more complete model of emission, all information about the size of \el~ is surrogate information and can never be interpreted asteroseismologically.
\clearpage
\clearpage

Fitting the other pulsational parameters -- amplitudes, periods, and phases -- requires that we minimize an appropriate metric for the mismatch between the measured Stokes parameters and the model Stokes parameters.  This is complicated by the lack of circular polarization in our model but its presence in the star.  The measured circular polarization, while not as strong as the linear polarization, does contribute to the intensity in each pulse and may or may not follow the shape of the pulse profile.  Since we are only modeling the total and linear intensity, we remove the measured $V$ from the measured $I$ by constructing the Stokes parameter $I_{new} = \sqrt{I^2-V^2}$ for each pulse. Otherwise we ignore the measured Stokes parameter $V$.  

A further complication arises from the appearance in the data of polarization vector changes.  In our model, the displacement and velocity polarization modes are orthogonal polarization states, which allows them to be described by a single Stokes parameter, e.g. $Q$, and leaves the other Stokes parameter zero.  \markcite{des01}{Deshpande} \& {Rankin} (2001) have shown that removal of the Radhakrishnan and Cooke vector rotation from the PSR B0943+10 data moves all of the subpulse behavior to one Stokes parameter, which they call $Q'$, and leaves only noise in the Stokes parameter $U'$, further confirming that the variations of PSR B0943+10 come from the superposition of orthogonal polarization modes.  We have adopted the notation of \markcite{des01}{Deshpande} \& {Rankin} (2001) and refer to the Stokes parameters calculated by our model before the incorporation of any vector rotation as primed quantities.  This gives us the option to conduct our fits in either the observed Stokes parameters (unprimed space) or in the frame of the star where the theoretical values are calculated (primed space).  We have chosen to do the fitting in observational (unprimed) space because it does not require any further alteration to the data.  All the fits we will present are the result of the simultaneous minimization of $I_{model} - I_{new}$, $Q_{model} - Q_{data}$, and $U_{model} - U_{data}$.  The quantities $I_{model}$, $Q_{model}$ and $U_{model}$ are calculated using Equations \ref{eqn:I}, \ref{eqn:Qmodel}, and \ref{eqn:Umodel}.

The final complication, and the most vexing, is the large, apparently-stochastic, pulse height variations.  To minimize the effects of the pulse height variation on our fits, we have normalized the data to remove the pulse-to-pulse amplitude variations.  To do this, we constructed an average pulse shape from the entire data set and normalized it to unit amplitude.  We then normalized each individual pulse in the Stokes parameter $I$ by multiplying each pulse by a single scale factor.  The scale factor was based upon the maximum intensity in each individual pulse.  This maximum does not always occur at the center of the pulse profile, so we calculated the scale factor for each pulse from the ratio of its maximum to the average pulse at the same longitude.  We treated the linear polarization parameters in a similar manner by constructing the linear polarization ($L = \sqrt{Q^2+U^2}$).  We normalized this quantity to unit amplitude in the same way that was done for $I$.  Then we scaled the individual pulses in $Q$ and $U$ by the maximum of the linear polarization for that pulse, effectively reducing both $Q$ and $U$ by the same normalization factor.   This process leaves the ratio of $Q$ to $U$ the same, thus having no effect on the value of the polarization angle.  Since the instrumental noise is normalized in the same manner as the rest of the pulse, an amplitude modulation can still be present as the ratio of pulse height to noise is preserved.  However, the normalization process reduces any amplitude modulations while still retaining the structure in the individual pulses. 

Normalization contained in the pulse-to-pulse variations makes the fitting exercise tractable but removes any information contained in the amplitude variations.  Before this normalization process, fits to the data did not consistently converge to solutions.  Afterward, the fits are always well-behaved but can no longer yield any astrophysical information about pulse amplitudes.  However, frequency and phase are only minimally affected.  We have demonstrated this by recalculating and examining all of the figures in \S\ref{data} for the normalized data.  The appearance of the driftbands and longitude and harmonic fluctuation spectrum remain essentially unchanged.

\subsection{\textit{Fits to the Polarization Angle}}
\label{FitstoPA}

Fits to the polarization angle swing, depicted in Figure \ref{fig:PulseProfile}, require four parameters: \al, \bt, \phio~ (the rotational longitude of the magnetic pole), and \chio~ (the position angle of the linear polarization at \phio).  As we have already mentioned, these parameters cannot be fit independently.  For any given value of \al~ and \phio~, a corresponding \bt~ and \chio~ can be fitted, and there is no substantial difference in the quality of the fits.  However, only one value of the \textit{ratio} of \al~ to \bt~ yields good fits and we have preserved the best ratio in all of our work.  The fits showed \phio~ and \chio~ are also highly correlated.  

\begin{figure}
\begin{center}
\includegraphics{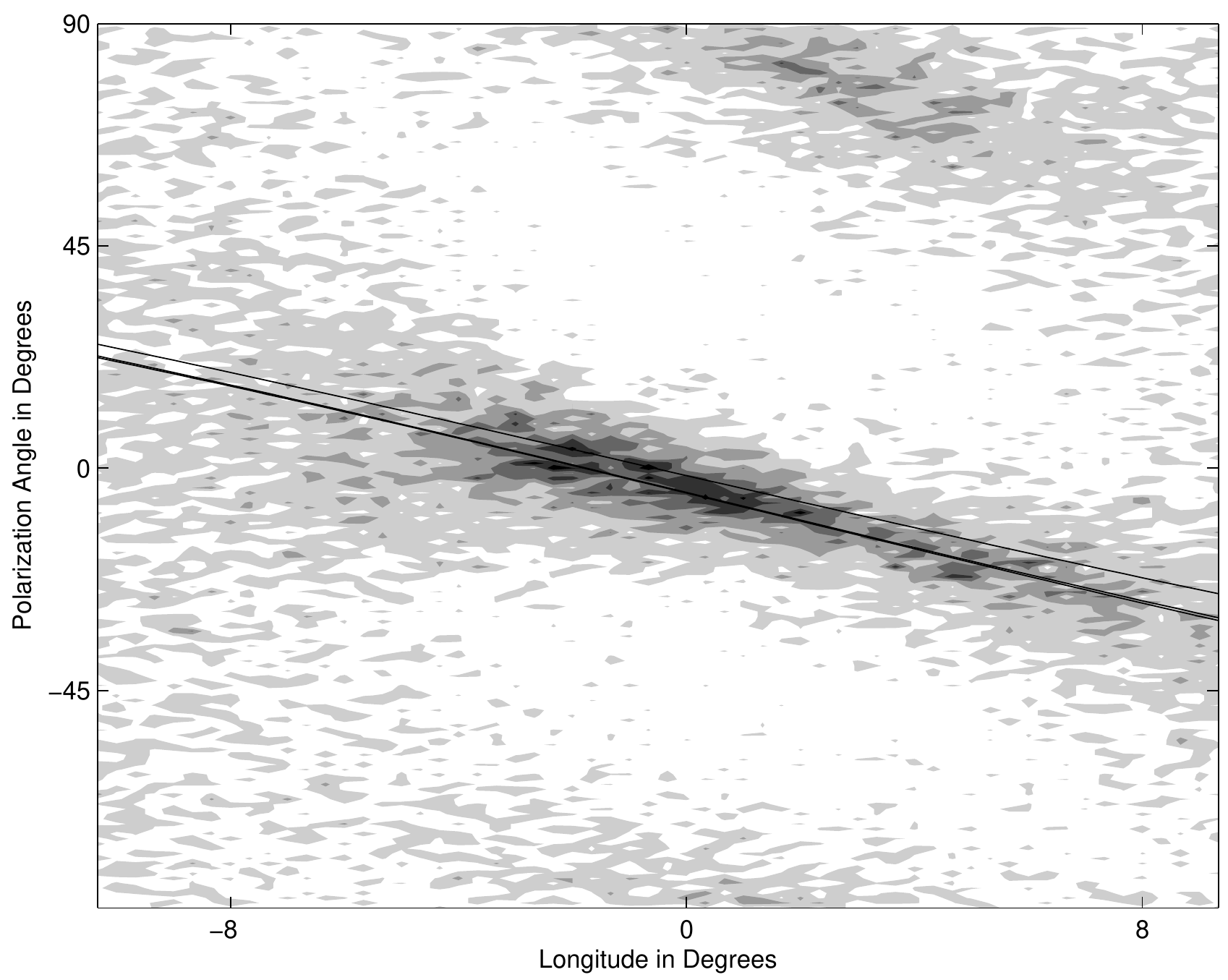}
\end{center}
\caption{The polarization angle histogram of all 816 pulses of PSR B0943+10.  The lines represent different values of \al, \bt, \phio, and \chio.  The top line is the polarization angle calculated from the values determined by \markcite{des01}{Deshpande} \& {Rankin} (2001), given in Table \ref{table:angle}.  The bottom line is the result of Fits 1 and 2 in Table \ref{table:angle}, which are indistinguishable.}
\label{fig:angle}
\end{figure}

Figure \ref{fig:angle} shows the \markcite{rad69}{Radhakrishnan} \& {Cooke} (1969) polarization angle calculated from Equation \ref{eqn:chimodel} superimposed on the polarization angle histogram of the data using the values shown in Table \ref{table:angle}.  These values were constrained to keep \al~ near the values given in \markcite{des01}{Deshpande} \& {Rankin} (2001).  Our fit has a different slope than that of \markcite{des01}{Deshpande} \& {Rankin} (2001) which cannot be matched by only adjusting \phio.  Inspection of Figure \ref{fig:angle} shows that our fit is slightly better than that of \markcite{des01}{Deshpande} \& {Rankin} (2001). 

\clearpage

\begin{table}
\begin{tabular}{cc|ccc}
Fit 1 & Fit 2 & DR2001 \\
\hline
\bt = $-2.81^\circ$ & \bt = $-5.42^\circ$ & \bt = $-4.29^\circ$ \\
\al = $8.03^\circ$ & \al = $15.51^\circ$ & \al = $11.58^\circ$ \\
\phio = $1.45^\circ$ & \phio = $0.45^\circ$ & \nodata \\
\chio = $-9.15^\circ$ & \chio = $-6.15^\circ$ & \nodata \\
\end{tabular}
\caption{The first two columns are the values of the geometrical parameters for two fits using Gaussfit (Fit 1 and Fit 2).  The last column contains the values used by \markcite{des01}{Deshpande} \& {Rankin} (2001) for \al~ and \bt.\label{table:angle}}
\end{table}

\clearpage

\clearpage

\subsection{\textit{Fits to the Pulsation Model}}
\label{FitstoPM}

Once we have settled upon values for parameters \al~ and \bt~ from geometric arguments, it is straightforward to fit our pulsational model to the data, but we must first choose a value for \el.  As we have already discussed, the features we observe in pulsar beams are spatially magnified, so that value of \el~ we use is a surrogate for a larger value at the stellar surface.  We can constrain the apparent \el~ by constraining the allowed pulsar geometry in the model.  As presented by \markcite{des01}{Deshpande} \& {Rankin} (2001), the subpulses in PSR B0943+10 drift all the way across the pulse window without large phase jumps or even driftband curvature, which is consistent with the identification of this pulsar as a conal signal pulsar \markcite{ran90}({Rankin} 1990).  In our model, where the subpulses are generated by a single frequency oscillation, continuous drift implies that our sightline does not cross a nodal line.  If we only allow geometries in which our sightline passes through one of the first three nodal regions, as plotted in Figure \ref{fig:beta}, and if we further exclude the possibility that our sightline passes through the central region of the star (which would be analogous to the core type classification of \markcite{ran90}{Rankin} (1990)), then \el~ is constrained by the absence of nodal lines in the average profile.  For this geometry, the lower and upper bounds on \el~ are 55 and 125, respectively.  We have chosen \el~ to be in the mid-range of these bounds, 75, in all the fits that follow.  In addition to fixing \el, we fix \Pone~ to be the known period of the pulsar.  We choose \phio, the closest approach to the magnetic pole, to be near the center of the average pulse profile based on the observations of the pulse profile of PSR B0943+10 at other frequencies \markcite{smi06}({Smits} {et~al.} 2006).

Once \al, \bt, \el, and \Pone~ are fixed, we use the following procedure to fit the remaining parameters:

\begin{itemize}
\item Choose trial values for all the parameters based on observed phases and amplitude ratios.
\item Generate a time series in $I$, $Q$, and $U$ in Equations \ref{eqn:I}, \ref{eqn:Qmodel}, and \ref{eqn:Umodel}, sampled in the same way as the data.
\item Impose a Gaussian window function on each pulse where the function is $g = e^{-(\phi-\phi_{mean})^2/(2\sigma)^2}$, where $\sigma \equiv 3.25^{\circ}$ and $\phi_{mean} \equiv 0^{\circ}$.
\item Calculate the metrics $I_{model} - I_{new}$, $Q_{model} - Q_{data}$, and $U_{model} - U_{data}$ and repeat until they are minimized.  
\end{itemize}

Because our Fourier analysis in \S\ref{qsft} showed that the frequency and phase of the subpulse wandered slightly over the 816 pulses, we fit the data in 100-pulse segments.  We implement the procedure above using Gaussfit.  Like our analysis of the multiple peaks near the subpulse frequency in \S\ref{qsft}, the variance in the data is much higher than the off-pulse instrumental noise, so we use the variance as an error estimate.  This artificially forces the reduced $\chi^2$ to be near one, so we have no absolute measure of the goodness of our fits nor will it ever be possible to do better for this pulsar unless we understand the physics behind the pulse height distribution.  

One other complication in our fit is that our model includes no unpolarized component.  This does not mean that our model radiation remains fully polarized, but rather that any unpolarized component arises from the superposition of the orthogonal polarization modes, i.e. if $Q_{model}$ and $U_{model}$ are equal, we have complete depolarization.  \markcite{des01}{Deshpande} \& {Rankin} (2001) have shown, and we concur, that PSR B0943+10 contains a substantial additional unpolarized component and the absence of this component from our model makes amplitudes we fit a compromise with reality. 

The best results of this fitting process are summarized in Table \ref{table:gaussfit}.  To be certain the fitted parameters are not dependent on how our sightline slices through the nodal region, we repeated these fits for all the values of \bt~ shown in Figure \ref{fig:beta}.  We performed a similar experiment by varying \phio.  Except for the expected $180^{\circ}$ difference in phase between the odd and even nodal regions, the results were the same as we have tabulated, indicating our fit is robust against changes in geometrical parameters.  Because our $\chi^2$ is only useful for guiding us to the best fit and is not able to tell us the likelihood that the data are consistent with our model, we cannot claim to have confirmed or ruled out a pulsational model.  The value of a quantitative fit is that it allows us to attempt to reproduce the data in an objective way rather than the subjective attempts in \markcite{cle04}{Clemens} \& {Rosen} (2004).  In the next section we will show the results of the simulations.  Fits to the remaining 100-pulse segments were not substantially different from the ones shown here.

We find that the fitted parameters can be divided into two categories: pulsational and geometrical parameters.  The pulsational parameters (e.g. the pulsational period, \Ptwo, and the spherical harmonic, $\Psi_{l,m=0}$) describe the surface variations that are centered on the magnetic pole.  The geometrical parameters (e.g. the offset between the rotation and magnetic axes, \al, and the distance between the magnetic pole and our line of sight, \bt) are highly correlated and it is their ratios rather than their precise values that are significant in matching the observations.  Therefore, our fits do not yield the full geometry of the star and external information has to be applied to know, for example, the inclination of the rotation axis.  Fortunately, the geometrical parameters are mostly independent from the pulsational parameters and thus the degeneracy of the geometric parameters does not prevent us from constraining the pulsational properties, a prerequisite for learning about the physics of pulsating neutron stars.  The simple average pulse profile and single frequency of PSR B0943+10 are reflected in a single set of pulsational parameters that would correspond in our model to a single pulsation mode.  In the more complex pulsars our model may demand multiple oscillation modes whose pulsational parameters would yield physical properties of neutron stars. 
 
\clearpage
\begin{table}
\begin{tabular}{c|ccc}
Parameter & Value & $\sigma$\\
\hline
\Pone & 1.097608 seconds & \nodata \\
\Ptwo & 0.031782 seconds &   6.07558e-08 seconds \\ 
\Pthree & 2.05337 seconds & \nodata \\   
\bt & $-2.81^\circ$ & \nodata \\
\al & $8.03^\circ$ & \nodata \\
\phio & $2^\circ$ & \nodata \\
\chio & $-9.15^\circ$ & $0.24^{\circ}$ \\
$l$ & 75 & \nodata \\
$a_{0,DPM}$ & 0.164 & 0.013 \\
$a_{0,VPM}$ & 0.1312 & 0.0027 \\
$a_{1,DPM}$ & 1.212 & 0.032 \\
$\psi_0$ & $-118.4^{\circ}$ & $2.4^{\circ}$ \\
$\psi_{delay}$ &  $53.9^{\circ}$ & $2.1^{\circ}$ \\
$\chi^2$ & 0.9338 & \nodata \\
\end{tabular}
\caption{The free and fixed parameters used in our model.  If the parameter was a free parameter, the value of $\sigma$ is given.  The results are from a fit of the first 100 pulses where \bt~ and \phio~ are fixed at $-2.8^{\circ}$ and $1.45^{\circ}$, respectively.  We get the values of \Pone~ from the Fourier transform of the entire run.  We calculate \Pthree~ from equation four in \markcite{cle04}{Clemens} \& {Rosen} (2004) such that $1/P_3 = 1/P_{time} - n/P_1$.  Fits to the other 100-pulse segments were not significantly different.\label{table:gaussfit}}
\end{table}

\clearpage

\section{Synthetic Pulsar Light Curves}
\label{synthetic}

We will now use our pulsational model and the best parameters fitted to the first 100 pulses of PSR B0943+10 to generate and analyze synthetic pulsar light curves.  In this way, we will learn whether our model has captured the essence of the variations in PSR B0943+10.  

\begin{figure}
\begin{center}
\includegraphics{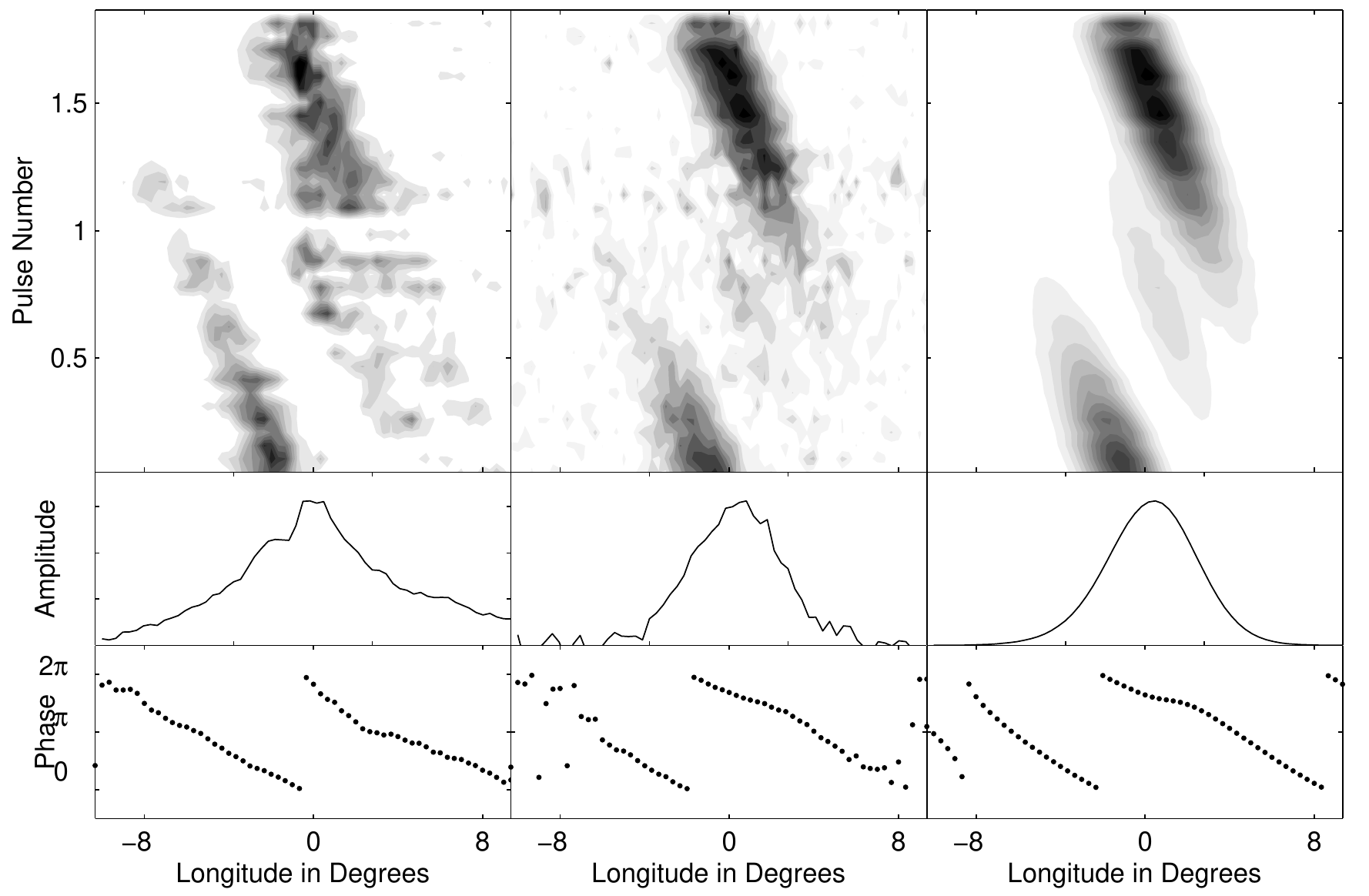}
\end{center}
\caption{Left panel: The first 100 pulses of PSR B0943+10 folded at \Pthree~ = 1.8708\Pone~ seconds, using the fitted value of \Pthree, compared to the left panel of Figure \ref{fig:driftdata}, where the data are folded at a value of \Pthree~ calculated from the Fourier transform.  Middle panel: Our model of the data, using the values of the parameters in Table \ref{table:gaussfit} calculated from Gaussfit.  We have added random noise to the simulation so the variance in the simulation approximately matches the variance in the data.  Right panel: Our same simulation, without noise.  The phase in the bottom panels were calculated using all 816 pulses in both the data and simulations.}
\label{fig:driftcomp}
\end{figure}

As in \S\ref{data}, we begin with the driftband.  Figure \ref{fig:driftcomp} shows a comparison between the first 100 normalized pulses of the data (left panel) and synthetic lightcurves of PSR B0943+10.  In the middle panel we have added random noise to the model to increase its variance to be approximately that of the data.  Our driftband shows the splitting on the right-hand-side of the profile as observed in the star.  In the model, it comes about because the time-like maxima in the velocity polarization mode are offset from the displacement polarization mode by an amount related to \psidelay.  Note that this means that the parallel track has a polarization $90^{\circ}$ different from the displacement polarization track, an effect reproduced in our polarization angle histogram which we discuss later. 

In our model, the amplitudes and the spatial envelopes of the velocity and displacement polarization modes are also different, because $\Psi_{l,m=0}$ has its maxima at antinodes and ${{\frac{\partial{\Psi_{l,m=0}}}{\partial{\theta_{mag}}}}}$ has its maxima at nodes.  This is why the velocity polarization mode track does not extend through the whole driftband.  This track is also asymmetric, appearing on the right-hand-side of the driftband and not the left.  In our model, this asymmetry arises from an offset between a maximum in the pulse window, $\phi_{mean}$, and the longitude of the magnetic pole, \phio.  While the difference between $\phi_{mean}$ and \phio~ is not related to pulsation parameters, it may lead to interesting insight in the pulsar emission mechanism. 
\clearpage

\begin{figure}
\begin{center}
\includegraphics{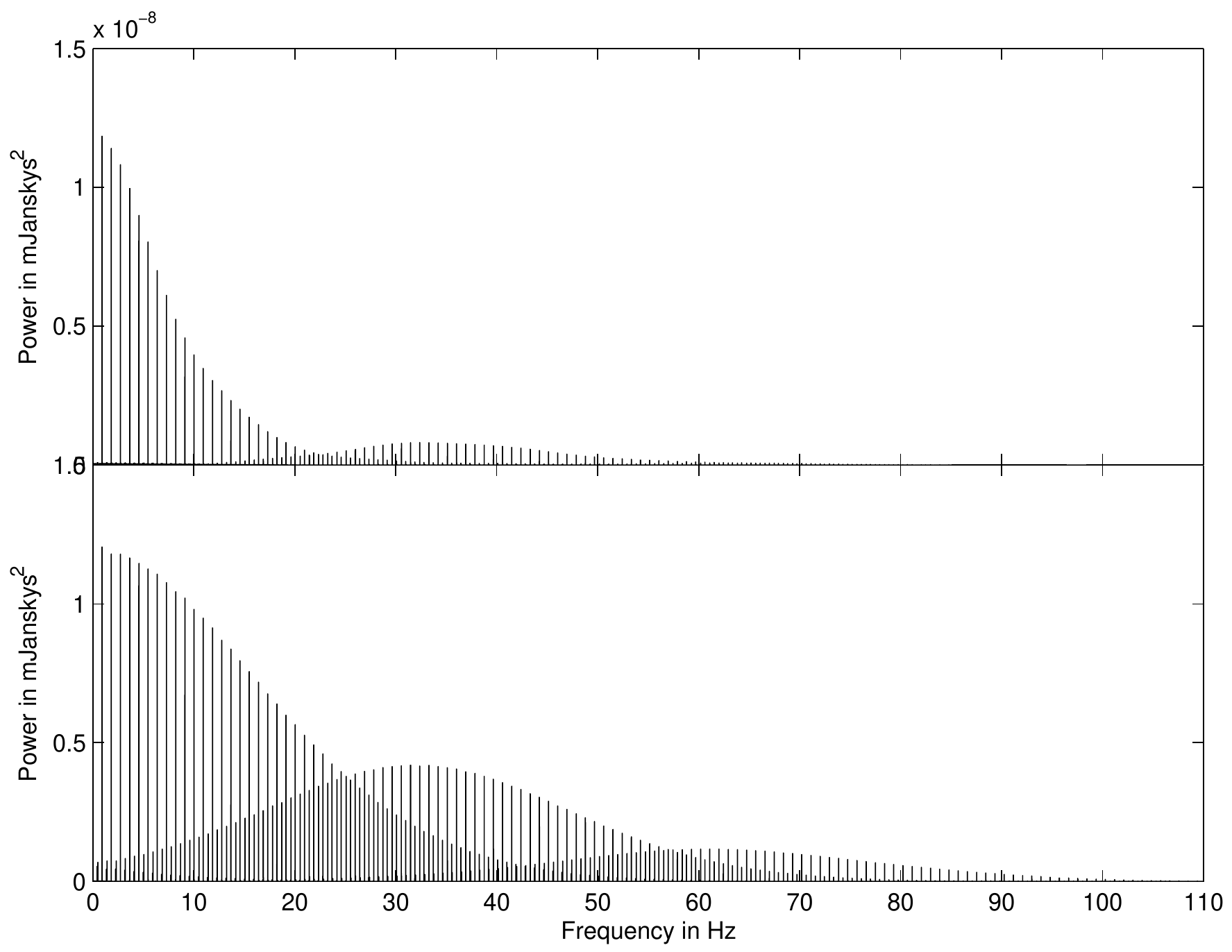}
\end{center}
\caption{Top panel: A fully-resolved Fourier transform all 816 normalized pulses from the archival data of PSR B0943+10.  Bottom panel: A Fourier transform of our simulated data where we have added random noise.  The peaks centered around 30 Hz are the subpulse frequency and its aliases.  The harmonic of the subpulse frequency and its aliases are centered around 65 Hz.  The values of the parameters used in our model are listed in Table \ref{table:gaussfit}.}
\label{fig:FTsim}
\end{figure}
\clearpage

We can also make a visual comparison of our pulsational model to the data in Fourier space.  We used the model parameters in Table \ref{table:gaussfit} to create 816 pulses matching the resolution of the data.  The bottom panel of Figure \ref{fig:FTsim} shows a fully-resolved Fourier transform of our simulated data where we have added random noise to our simulation.  The Fourier transform of our simulation does not show the splitting in the subpulse frequency, as discussed in \S\ref{data}, because we have created our model only using a single subpulse frequency.  To reproduce the closely spaced peaks, our model would require the introduction of additional, incommensurate subpulse frequency or the subpulse frequency would have to wander as a function of time.  For comparison, the top panel of Figure \ref{fig:FTsim} shows the fully-resolved Fourier transform of the normalized data.  

\begin{figure}
\begin{center}
\includegraphics{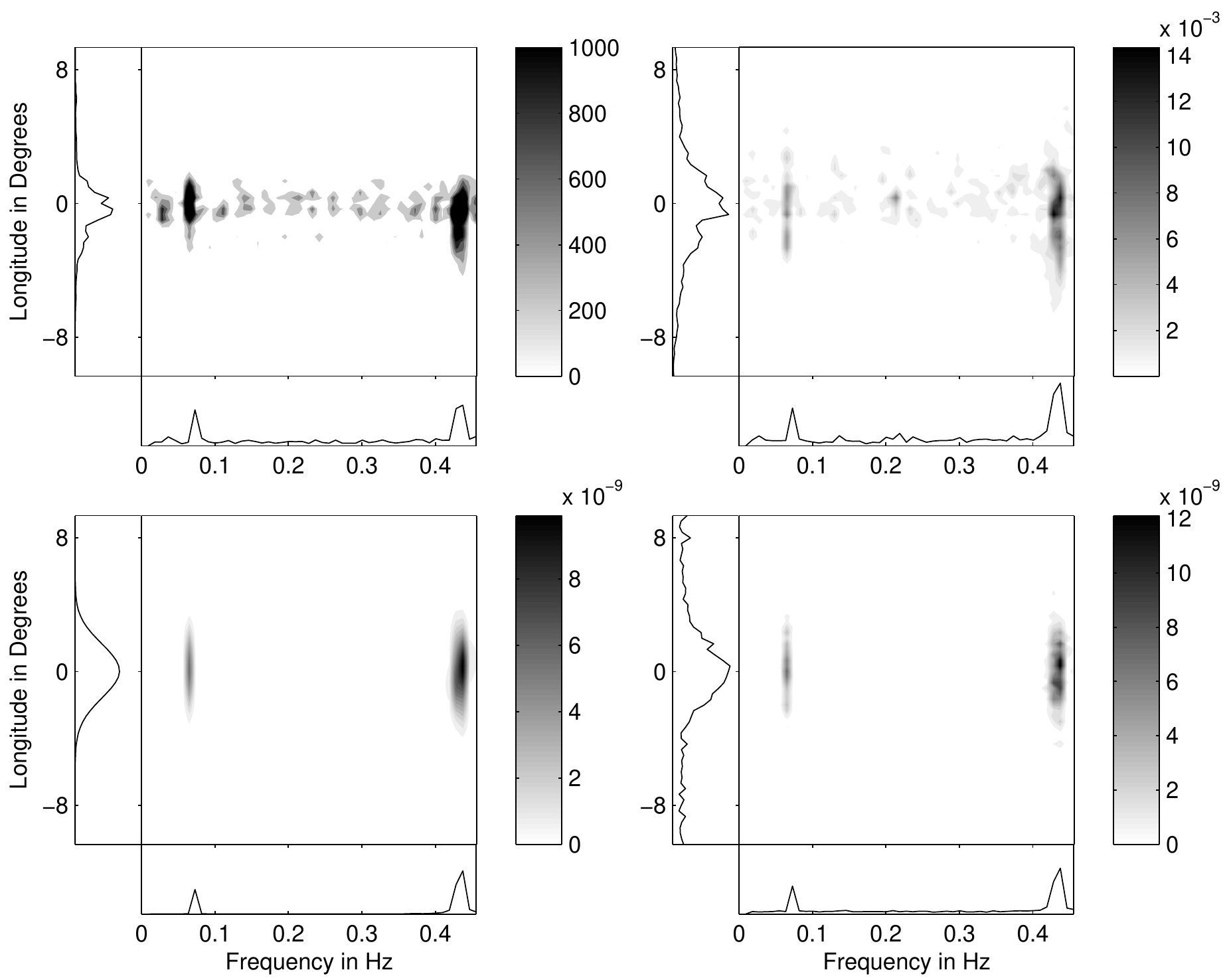}
\end{center}
\caption{Upper panels: The longitude resolved fluctuation spectra of the first 100 pulses of data.  The left-hand-side shows the raw data while the right-hand-side shows the data after our normalization process.  Lower panels: The longitude resolved fluctuation spectra of 100 pulses of simulated data using the values of the parameters listed in Table \ref{table:gaussfit} (left) and with the addition of noise (right).  The contour levels are in mJy$^2$.}
\label{fig:LRFcompare}
\end{figure}

\begin{figure}
\begin{center}
\includegraphics{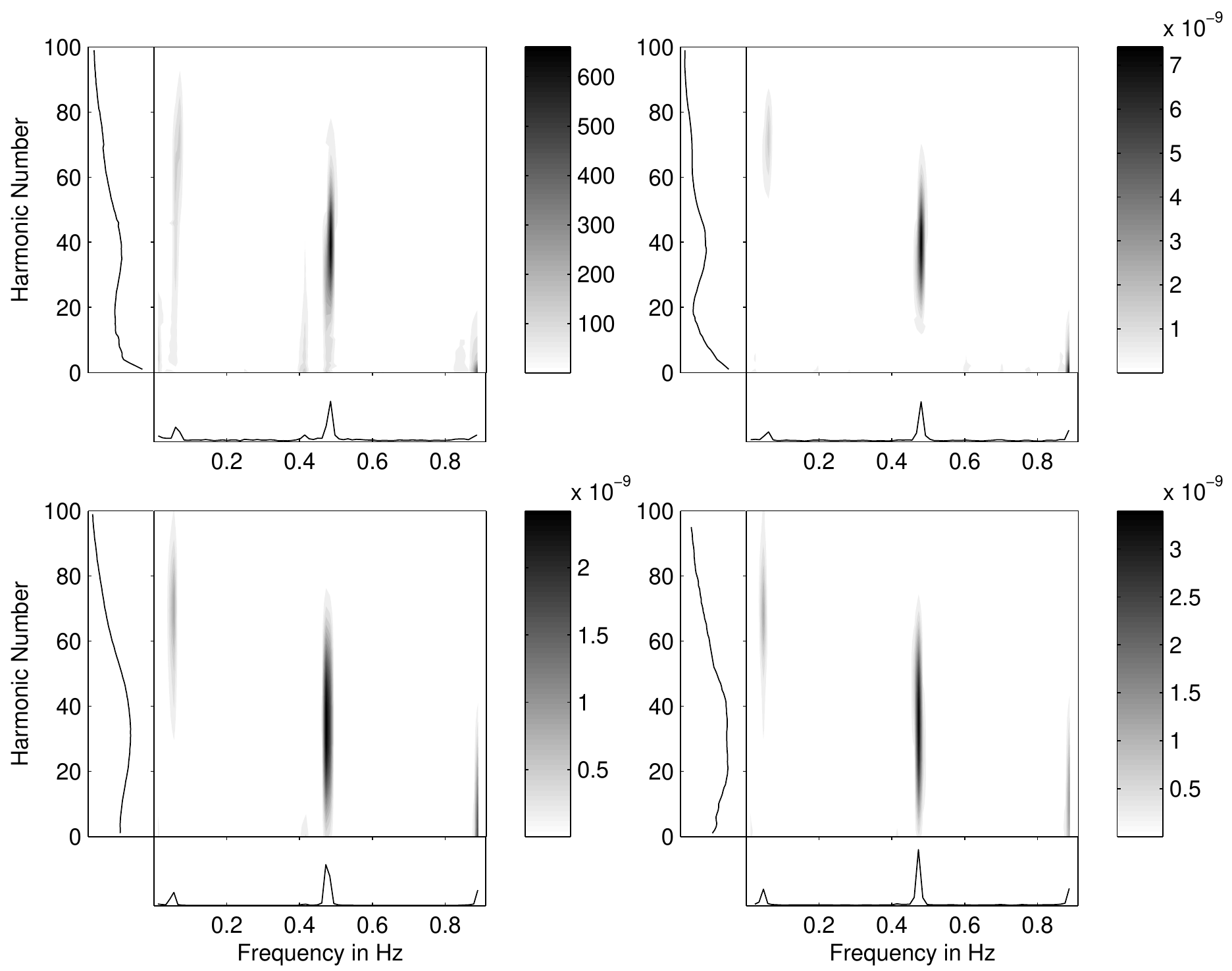}
\end{center}
\caption{Upper panels: The harmonic resolved fluctuation spectra of the first 100 pulses of data.  The left-hand-side shows the raw data while the right-hand-side shows the data after our normalization process.  Lower panels: The harmonic resolved fluctuation spectra of 100 pulses of simulated data using the values of the parameters listed in Table \ref{table:gaussfit} (left) and with the addition of noise (right).  The contour levels are in mJy$^2$.}
\label{fig:HRFcompare}
\end{figure}

While the unfolded Fourier transform does not reproduce the amplitude of the harmonic structure well, this is not surprising because we are modeling the subpulses, not the low frequency pulse window.  Despite this, both the harmonic and longitude-resolved fluctuation spectra, shown in Figures \ref{fig:LRFcompare} and \ref{fig:HRFcompare} respectively, closely reproduce the data.  In both figures, we show the spectra of first 100 pulses of data that were fit using Gaussfit, both unaltered and normalized.  The bottom panels show the fluctuation spectra from our simulation with and without noise.

 \clearpage

\begin{figure}
\begin{center}
\includegraphics{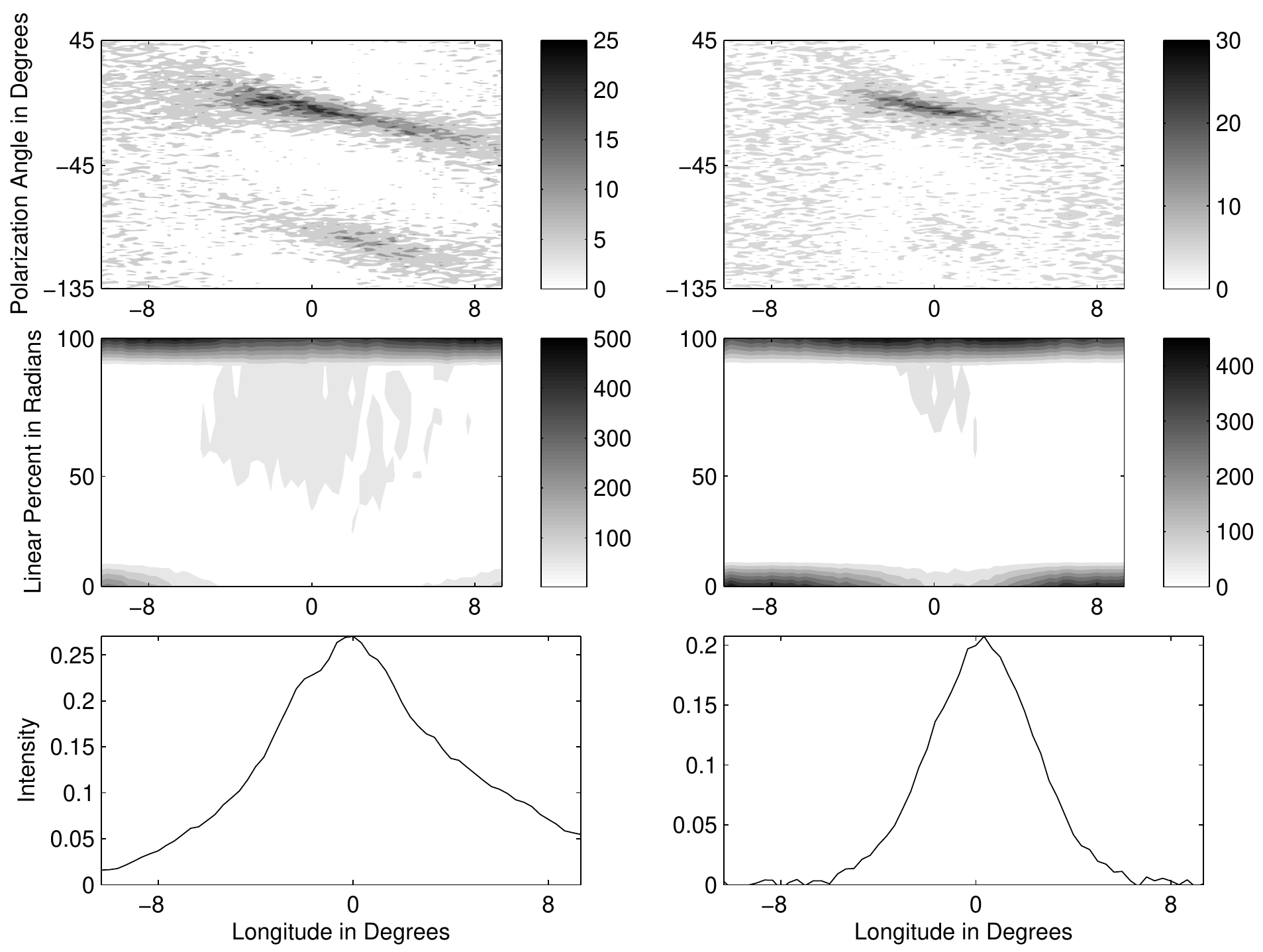}
\end{center}
\caption{Left panel: The properties of PSR B0943+10 using all 816 pulses in the archival data.  The top two panels show in histogram form the polarization angle and the fractional linear polarization percent.  The average pulse shape is in the bottom panel.  Right panel: Our model of the data, using the results of Gaussfit.  Noise was adding using the off-pulse noise from the archival data.}
\label{fig:stinebring}
\end{figure}
\clearpage

We can further compare our model to the data using plots similar to those in \markcite{sti84a,sti84b}{Stinebring} {et~al.} (1984a, 1984b).  We recreate plots of the polarization angle, polarization fraction, and total intensity.  Figure \ref{fig:stinebring} shows the normalized data of all 816 pulses in the left panel and all 816 pulses from our simulation in the right panel.  We have added noise to our simulation in the same way as we did in the driftband plot in Figure \ref{fig:driftcomp}.  Our model pulse shape reflects the Gaussian window imposed and does not match the data in the wings of the profile.  Our polarization fraction is more tightly grouped at 0\% and 100\%, in part because we do not include an unpolarized component in our model, as we discussed in \S\ref{data}.  Our velocity polarization mode is weaker than the corresponding track in the angle histogram of the data.  Nonetheless, Figure \ref{fig:stinebring} represents a dramatic improvement in our ability to model pulsar data.  It is based on an objective fit that was not constrained by the quantities in this plot but by the lightcurve directly.  The striking similarities between our model and the data give us confidence that our pulsation model can reproduce significant features in PSR B0943+10.

\section{Summary and Conclusions}
\label{conc2}

Our objective in this paper has been first to fit single pulse data of PSR B0943+10 using a pulsational model and second, to create simulated pulsar lightcurves from the best fit.  The model we use was developed in \markcite{cle04}({Clemens} \& {Rosen} 2004) and in a companion paper to this one \markcite{cle07}({Clemens} \& {Rosen} 2007).  It is founded upon pulsational displacements and their associated surface velocities, which result in the emission of distinct polarization modes orthogonal to each other.

We chose PSR B0943+10 as an initial target for our study because it has a simple morphology and is well-studied in the literature.  We were surprised in our analysis of published data to find undocumented behaviors in the form of driftband splitting, frequency splitting, and phase wandering of the subpulses.  The driftband splitting is a stable feature of the data while the frequency and phase wandering are more erratic and in need of further observational study.  The model we use is based on a single pulsational frequency and cannot reproduce the erratic behavior, but it has been able to reproduce the driftband splitting.  

Our fitting exercise has been successful; our least squares fitting algorithm converged to a solution for an amplitude normalized data set from which we had removed circular polarization.  The fitted parameters for this converged model represent an objective application of our pulsation model to PSR B0943+10.  We have used them to create synthetic data that reproduce the stable features of the observations, including the pulse shape, the polarization angles, and the orthogonal mode morphology.  Our model cannot reproduce the pulse amplitude distribution and therefore we cannot quantitatively access the validity of our model beyond its ability to reproduce the more stable features of the data.

If we accept a pulsational model for the drifting subpulses in PSR B0943+10, then some of our fitted parameters have physical significance.  \Ptwo~ represents the eigenperiod of a single pulsation mode.  The spherical harmonic degree of this mode, \el, is unknown for reasons we have explained.  The amplitudes we measured and the value of \el~ are not meaningful without better understanding of the radio emission mechanism.  The fitted quantity \psidelay~ reflects an adjustment between the phase of the pulsational displacements and pulsational velocities, which would be $90^{\circ}$ for an adiabatic pulsation.  Positive values of \psidelay~ mean that maximum heating follows maximum compression and requires some mechanism for delaying the flux changes.  Curiously, the sign and value of \psidelay~ in our fits to PSR B0943+10 are similar to the measured value for pulsating white dwarfs in which the flux is delayed by a surface convection zone \markcite{gol99}({Goldreich} \& {Wu} 1999).  Our value of \psidelay~ is consistent with the identification of the displacement polarization mode as a surface flux phenomena related to pulsations. 

From an astroseismological perspective, PSR B0943+10 is not the most interesting target for study.  The simplicity that makes it attractive for testing our model limits the useful physics that can be extracted.  There are other pulsars, e.g. PSR B0031-07, that show multiple subpulse frequencies and others that appear to show correlated oscillations between the magnetic poles \markcite{wel07}({Weltevrede}, {Wright}, \&  {Stappers} 2007).  It will be an interesting challenge to apply our pulsational model to objects like these. \markcite{edw06}{Edwards} (2006) has proposed the phase behavior of subpulses as a critical test of the drifting spark model and his argument applies equally well to our model.  It is possible that broader studies of more complex pulsars will be able to rule out non-radial oscillations as the origin of their subpulses, but for now, PSR B0943+10 remains an encouraging example of a pulsar whose subpulse behavior can be reproduced by stellar pulsation model.

\bibliography{}
\end{document}